\documentclass[aps,pra,reprint,twocolumn,superscriptaddress,floatfix,nofootinbib]{revtex4-1}
\usepackage{graphicx,amsmath,amsfonts,amssymb,amsthm,xr}
\usepackage{epsfig,amsmath,amssymb,color,dsfont,upgreek,physics}
\usepackage{mathrsfs}
\usepackage{mathtools}
\usepackage{bbold}

\usepackage[bookmarks=true,colorlinks,linkcolor=OrangeRed,urlcolor=NavyBlue,citecolor=RoyalBlue]{hyperref}
\usepackage[dvipsnames]{xcolor}

\definecolor{mygold}{rgb}{0.93,0.69,0.13}

\definecolor{mypurple}{rgb}{0.49,0.18,0.56}

\definecolor{mygreen}{rgb}{0,0.5,0}

\definecolor{mygreen}{rgb}{0,0.5,0}
\definecolor{myred}{rgb}{0.7,0,0}
\usepackage[normalem]{ulem}

\begin{document}
\title{Suppressing nonperturbative gauge errors in the thermodynamic limit using local pseudogenerators}
\author{Maarten Van Damme}
\thanks{These authors contributed equally to this work.}
\affiliation{Department of Physics and Astronomy, University of Ghent, Krijgslaan 281, 9000 Gent, Belgium}
\author{Julius Mildenberger}
\thanks{These authors contributed equally to this work.}
\affiliation{INO-CNR BEC Center and Department of Physics, University of Trento, Via Sommarive 14, I-38123 Trento, Italy}
\author{Fabian Grusdt}
\affiliation{Department of Physics and Arnold Sommerfeld Center for Theoretical Physics (ASC), Ludwig-Maximilians-Universit\"at M\"unchen, Theresienstra\ss e 37, D-80333 M\"unchen, Germany}
\affiliation{Munich Center for Quantum Science and Technology (MCQST), Schellingstra\ss e 4, D-80799 M\"unchen, Germany}
\author{Philipp Hauke}
\affiliation{INO-CNR BEC Center and Department of Physics, University of Trento, Via Sommarive 14, I-38123 Trento, Italy}
\author{Jad C.~Halimeh}
\email{jad.halimeh@physik.lmu.de}
\affiliation{INO-CNR BEC Center and Department of Physics, University of Trento, Via Sommarive 14, I-38123 Trento, Italy}

\begin{abstract}
With recent progress in quantum simulations of lattice-gauge theories, it is becoming a pressing question how to reliably protect the gauge symmetry that defines such models. In a recent work [J.~C.~Halimeh \textit{et al.}, \href{https://arxiv.org/abs/2108.02203}{arXiv:2108.02203}], an experimentally feasible gauge-protection scheme has been proposed that is based on the concept of a \textit{local pseudogenerator}, which is required to act identically to the full gauge-symmetry generator in the target gauge sector, but not necessarily outside of it. The scheme has been analytically and numerically shown to reliably stabilize lattice gauge theories in the presence of perturbative errors on finite-size analog quantum-simulation devices. In this work, through uniform matrix product state calculations, we demonstrate the efficacy of this scheme for nonperturbative errors in analog quantum simulators up to all accessible evolution times in the thermodynamic limit, where it is \textit{a priori} neither established nor expected that this scheme will succeed. Our results indicate the presence of an emergent gauge symmetry in an adjusted gauge theory even in the thermodynamic limit, which is beyond our analytic predictions. Additionally, we show through quantum circuit model calculations that gauge protection with local pseudogenerators also successfully suppresses gauge violations on finite quantum computers that discretize time through Trotterization. Our results firm up the robustness and feasibility of the local pseudogenerator as a viable tool for enforcing gauge invariance in modern quantum simulators and NISQ devices.
\end{abstract}

\date{\today}
\maketitle
\tableofcontents

\section{Introduction}
Gauge theories are the most complete framework for the description of elementary particles and their interactions as mediated by gauge bosons \cite{Weinberg_book}. A gauge theory's principal property is local gauge symmetry, which encodes physical laws of nature through local constraints in both space and time \cite{Zee_book}. A paradigmatic example is Gauss's law in quantum electrodynamics, which imposes an intrinsic relation between the distribution of charged matter and the surrounding electromagnetic field \cite{Gattringer_book}. Examples of physical features arising from such local constraints include the long-ranged Coulomb law and massless photons \cite{Weinberg_book}. 

The simulation of gauge theories on classical computers remains a challenging task due to the exponentially large Hilbert spaces involved. Given their importance in modern physics, an accessible means of investigating their physics is highly desirable. Over the past few years, there has been a significant experimental drive to realize gauge theories in quantum synthetic matter (QSM) setups \cite{Martinez2016,Muschik2017,Bernien2017,Klco2018,Kokail2019,Goerg2019,Schweizer2019,Mil2020,Klco2020,Yang2020,Zhou2021}. On the one hand, this is due to the impressive advancement in the level of precision and control in such setups \cite{Bloch2008}, which naturally makes the quantum simulation of gauge theories a realistic possibility \cite{Wiese_review,Zohar_review,Dalmonte_review,Pasquans_review,Alexeev_review,aidelsburger2021cold,zohar2021quantum,klco2021standard}. On the other hand, this facilitates an easily accessible table-top framework for addressing fundamental questions from the fields of high-energy physics and strongly correlated electrons. For example, the thermalization of gauge theories and how it connects to early-time far-from-equilibrium dynamics, as e.g.~investigated in violent collisions of heavy nuclei, is not fully settled \cite{Berges_review}. Recently, the thermalization dynamics in a large-scale $(1+1)-$dimensional $\mathrm{U}(1)$ lattice gauge theory (LGT) has been demonstrated on a 71-site tilted Bose--Hubbard superlattice \cite{Zhou2021}, which offers a way forward in understanding this connection.
 
Stabilizing gauge symmetries in QSM implementations of LGTs is essential for the faithful modeling of their physics. Indeed, even though gauge invariance is a postulate in nature, it does not lend itself directly to such experimental realizations, with the exception of special realizations that integrate out either gauge or matter fields \cite{Martinez2016,Bernien2017,Kokail2019}. Without indefinite fine-tuning, implementations of LGTs with dynamical matter and gauge fields as active degrees of freedom will inevitably suffer from errors that break gauge symmetry \cite{Halimeh2020a}. The gauge symmetry has to be engineered directly in the LGT realization, and this can be a daunting task. A lot of theoretical work has been directed at enforcing gauge invariance in QSM setups of LGTs through energetic constraints \cite{Zohar2011,Zohar2012,Banerjee2012,Zohar2013,Hauke2013,Stannigel2014,Kuehn2014,Kuno2015,Yang2016,Kuno2017,Negretti2017,Dutta2017,Barros2019,Lamm2020,Halimeh2020e,Kasper2021nonabelian,Halimeh2021gauge,Halimeh2021stabilizing}. However, most gauge-protection schemes based on energy-penalty constraints originally comprised adding a term to the Hamiltonian that is proportional to the square of the gauge-symmetry generator. Even though this is equivalent to turning the target sector into a ground-state manifold and thereby reliably suppressing processes that lead away from it, the approach is in general experimentally challenging as it involves multibody terms that can be significantly harder to engineer than the ideal gauge theory itself. This poses a principal challenge since without \textit{controlled} gauge violations in an experiment, one cannot firmly establish the fidelity of the observed gauge-theory physics.

Recently, an experimentally friendly scheme has been proposed for the $(1+1)-$dimensional $\mathrm{U}(1)$ quantum link model (QLM) \cite{Wiese_review,Chandrasekharan1997} based on weighted terms linear in the gauge-symmetry generator \cite{Halimeh2020e}. This constructs a protection term with only single-body terms, which are easier to implement than the ideal theory itself. This \textit{linear gauge protection} has been shown to work up to indefinite times on finite systems \cite{Halimeh2020e}, and up to all accessible times in the thermodynamic limit \cite{vandamme2021reliability}. The method has also been experimentally shown to reliably stabilize gauge invariance in an ultracold-atom implementation of the $\mathrm{U}(1)$ QLM \cite{Yang2020}. Extensions of the scheme to non-Abelian LGTs have also displayed promising results in exact diagonalization (ED) \cite{Kasper2021nonabelian,Halimeh2021gauge}. However, linear gauge protection brings forth no advantage for certain gauge theories such as the $\mathbb{Z}_2$ LGT \cite{Zohar2017,Borla2020}, whose gauge-symmetry generator is itself a multibody operator proportional to its own square.

\begin{figure}[t!]
	\centering
	\includegraphics[width=.48\textwidth]{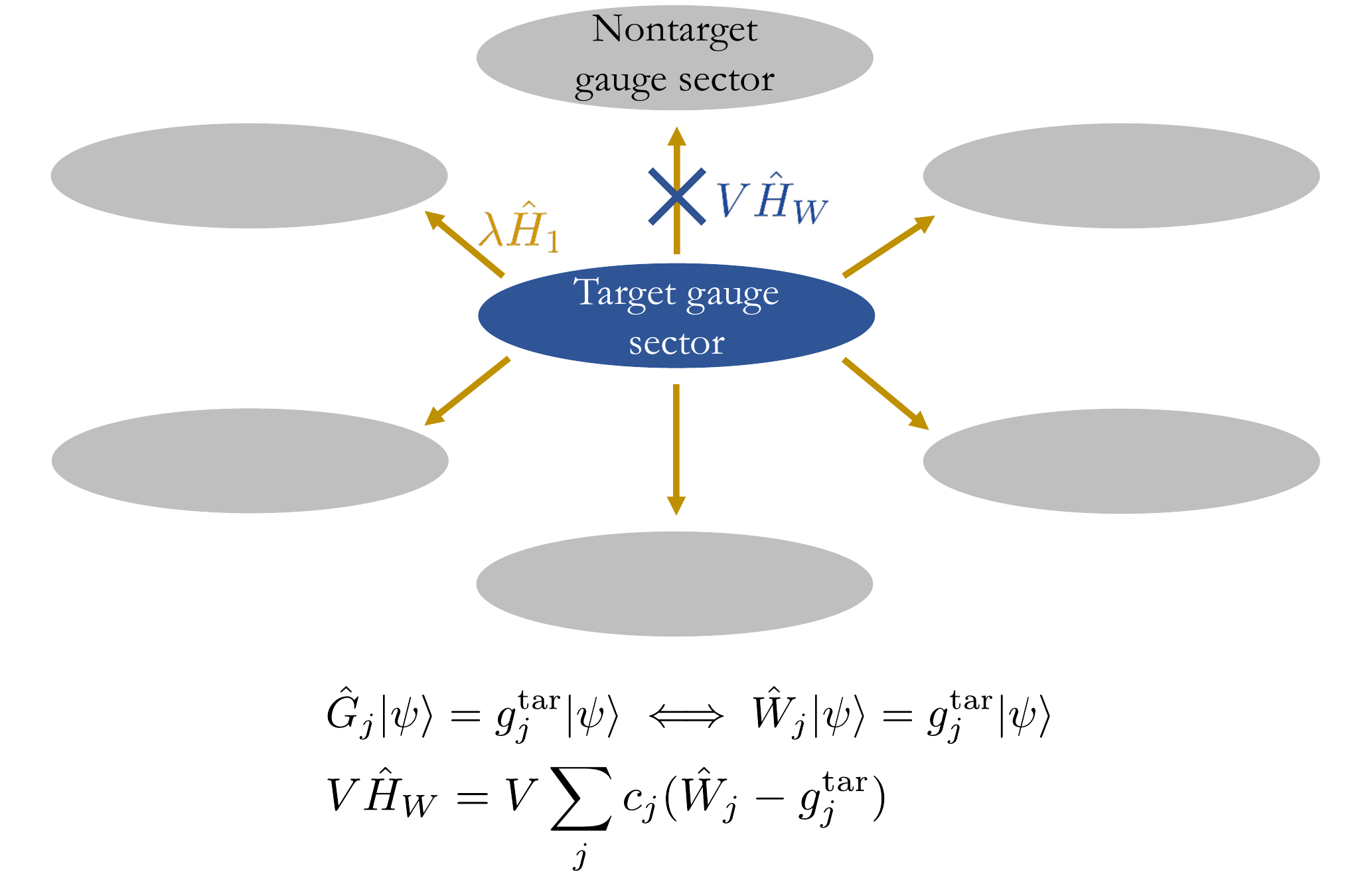}
	\caption{(Color online). Schematic illustrating the concept of local-pseudogenerator protection. The local pseudogenerator (LPG) $\hat{W}_j$ acts identically to the full generator $\hat{G}_j$ in the target gauge sector, but not necessarily outside of it. Starting in the target gauge sector $\mathbf{g}^\text{tar}=(g_1^\text{tar},g_2^\text{tar},\ldots,g_L^\text{tar})$, the ideal theory $\hat{H}_0$ propagates the dynamics within this sector, but experimentally unavoidable errors $\lambda \hat{H}_1$ create transitions to other gauge-invariant sectors (yellow). At sufficiently large but experimentally feasible values of the protection strength $V$, the LPG protection $V\hat{H}_W=V\sum_jc_j(\hat{W}_j-g_j^\text{tar})$, with $c_j$ a suitably chosen sequence (see main text for details), will energetically penalize these transitions. It will controllably suppress gauge violations in the thermodynamic limit for all accessible evolution times even when the error strength $\lambda$ is not perturbative.}
	\label{fig:schematic} 
\end{figure}

Nevertheless, the core concept from the theory of linear gauge protection has been extended to the case of the $\mathbb{Z}_2$ LGT by employing a \textit{local pseudogenerator} (LPG) that acts identically to the full gauge-symmetry generator in the target gauge sector, but not necessarily outside of it \cite{Halimeh2021stabilizing}; see Fig.~\ref{fig:schematic} and Sec.~\ref{sec:LPG}. This leaves the LPG with significantly fewer engineering requirements than its full counterpart, rendering it ideal for QSM implementations. LPG protection has been shown to work robustly for perturbative errors on finite-size quantum simulators up to all accessible evolution times in ED, and analytic derivations have demonstrated that the protected dynamics is faithfully reproduced by emergent gauge theories with well-defined timescales. However, for experimentally feasible implementations in the thermodynamic limit, the analytic formalism of Ref.~\cite{Halimeh2021stabilizing} does not guarantee reliable gauge protection, and even predicts its complete breakdown in a worst-case scenario. 

In this work, we test the performance of LPG protection on infinite-size analog quantum simulators in the presence of nonperturbative gauge-breaking errors using infinite matrix product state (iMPS) techniques \cite{mps2}, which work directly in the thermodynamic limit. Our numerical results demonstrate controlled gauge violations for sufficiently large volume-independent protection strengths. Moreover, we also test the LPG protection on a numerical simulation of a finite-size quantum computer that employs Trotterized dynamics, showing excellent performance for experimentally relevant settings.

The rest of our paper is organized as follows: In Sec.~\ref{sec:LPG}, we review the theory of LPG protection. In Sec.~\ref{sec:model}, we introduce the main model considered in this work, along with the quench protocol employed for the dynamics we calculate. Our numerical results are then presented and discussed in Secs.~\ref{sec:analog} and~\ref{sec:digital} for the cases of analog quantum simulators and digital quantum computers, respectively. We conclude and provide future outlook in Sec.~\ref{sec:conc}. Appendix~\ref{app:supp} provides supporting numerical results, including a discussion on the choice of LPG sequences.

\section{Local pseudogenerator}\label{sec:LPG}
A gauge theory $\hat{H}_0$ is invariant under a gauge symmetry with generator $\hat{G}_j$, defining a local constraint centered at the spatial degree of freedom $j$. Gauge invariance is encoded in the commutation relations $[\hat{H}_0,\hat{G}_j]=0,\,\forall j$. The gauge-symmetry generator $\hat{G}_j$ has eigenvalues $g_j$, and a given configuration $\mathbf{g}=(g_1,g_2,\ldots,g_L)$ defines a gauge-invariant sector, where $L$ is the number of local constraints. In a realistic implementation of gauge theories with dynamical matter and gauge fields, gauge-breaking errors $\lambda\hat{H}_1$ at strength $\lambda$ will always be present, where $[\hat{H}_1,\hat{G}_j]\neq0$, and so gauge invariance is no longer preserved in their presence. Controlling gauge violations due to such errors is crucial if the quantum simulation is to yield reliable gauge-theory physics.

The theory of gauge protection through a local pseudogenerator has been introduced in Ref.~\cite{Halimeh2021stabilizing}. It is based on the following idea: assume the system to be governed by a faulty gauge theory $\hat{H}=\hat{H}_0+\lambda\hat{H}_1+V\hat{H}_W$, where a protection term $V\hat{H}_W$ of strength $V$ has been added for the purpose of suppressing gauge violations. If the dynamics of the system is to be restricted to a specific target gauge-invariant sector $\mathbf{g}^\text{tar}=(g_1^\text{tar},g_2^\text{tar},\ldots,g_L^\text{tar})$, then it suffices to construct the protection term $\hat{H}_W$ using LPGs $\hat{W}_j$ that act identically to the full generators $\hat{G}_j$ in the target sector, but not necessarily outside of it (see Fig.~\ref{fig:schematic}). Naturally, this significantly relaxes implementational requirements as it can allow to realize $\hat{W}_j$ with fewer-body terms than $\hat{G}_j$, which is highly desirable in current QSM implementations that are at the scale of a few sites or qubits. In particular, the LPG protection term takes the form
\begin{align}\label{eq:LPGprotection}
V\hat{H}_W=V\sum_jc_j\big(\hat{W}_j-g_j^\text{tar}\big),
\end{align}
where $c_j$ are appropriately chosen real coefficients, as we will discuss in the following. At every local constraint defined by the spatial coordinate $j$, the LPG must satisfy the relation
\begin{align}
\hat{W}_j\ket{\psi}=g_j^\text{tar}\ket{\psi}\iff \hat{G}_j\ket{\psi}=g_j^\text{tar}\ket{\psi}.
\end{align}

Let us denote by $w_j$ the eigenvalues of $\hat{W}_j$. When the coefficients $c_j$ are rational numbers satisfying the \textit{compliance} condition 
\begin{align}\label{eq:compliance}
\sum_jc_j(w_j-g_j^\text{tar})=0\iff w_j=g_j^\text{tar},\,\forall j,
\end{align}
it can be analytically shown, by adapting the formalism of Ref.~\cite{abanin2017rigorous} for the slow heating of fast-driven systems, that gauge invariance is stabilized up to a timescale exponential in $V$ \cite{Halimeh2021stabilizing}. The power of the compliant sequence lies in the fact that it allows the LPG protection to energetically isolate the target sector from all other gauge-invariant sectors, such that any processes away from it become nonresonant at leading orders in $\lambda\hat{H}_1$; cf.~Fig.~\ref{fig:schematic}.

This \textit{compliant} sequence, however, will have to grow exponentially with system size in order to satisfy the compliance condition~\eqref{eq:compliance} at a given value of $V$. As such, it is impractical for large-scale implementations of LGTs, let alone in the thermodynamic limit. Turning to periodic \textit{noncompliant} sequences, one can still analytically prove and numerically demonstrate that LPG protection stabilizes gauge invariance up to a timescale $\tau_\text{adj}\propto V/(V_0L)^2$, where $V_0$ is a volume-independent energy term \cite{Halimeh2021stabilizing}. This is because LPG protection leads to the emergence of an adjusted gauge theory $\hat{H}_\text{adj}=\hat{H}_0+\lambda\hat{\mathcal{P}}_0\hat{H}_1\hat{\mathcal{P}}_0$, where $\hat{\mathcal{P}}_0$ is the projector onto the target gauge sector, up to an error upper bound $\propto tV_0^2L^2/V$, as can be derived using the quantum Zeno effect \cite{facchi2002quantum,facchi2004unification,facchi2009quantum,burgarth2019generalized}.

In the thermodynamic limit $L\to\infty$, it is clear that the above analytic arguments predict, in a worst-case scenario, a complete breakdown of LPG protection at any evolution time $t>0$, as then the error upper bound of the dynamics propagated by the emergent gauge theory $\hat{H}_\text{adj}$ will be $\propto tV_0^2L^2/V\to\infty$ for realistic values of $V$. However, recent works have shown that gauge invariance has intrinsic timescales over which it persists, even without protection, in the presence of perturbative errors \cite{Halimeh2020b,Halimeh2020c}. Furthermore, explicit gauge violations have been shown to remain localized even in large-scale simulators in the presence of linear protection \cite{Halimeh2020d,vandamme2021reliability}, performing better than the analytically predicted worst-case scenarios. Therefore, it is natural to ask not only whether LPG protection will fare well in the thermodynamic limit, but also whether it will do so in the presence of nonperturbative errors that can arise in QSM setups \cite{Schweizer2019}.

\section{Model and quench protocol}\label{sec:model}
We consider in this work the $(1+1)-$D $\mathbb{Z}_2$ LGT \cite{Zohar2017,Borla2020,Yang2020fragmentation,kebric2021confinement} with $L$ matter sites and $L$ gauge links, given by the Hamiltonian
\begin{align}\label{eq:H0}
\hat{H}_0=-J\sum_{j=1}^{L-1}\big(\hat{a}_j^\dagger\hat{\tau}^z_{j,j+1}\hat{a}_{j+1}+\text{H.c.}\big)-h\sum_{j=1}^L\hat{\tau}^x_{j,j+1}.
\end{align}
The hard-core bosonic annihilation (creation) operator is denoted by $\hat{a}_j^{(\dagger)}$ on matter site $j$, the number operator for hard-core bosons on matter site $j$ is $\hat{n}_j=\hat{a}^\dagger_j\hat{a}_j$, and the electric (gauge) field is represented by the Pauli matrix $\hat{\tau}^{x(z)}_{j,j+1}$ on the link between matter sites $j$ and $j+1$. The $\mathbb{Z}_2$ gauge symmetry of this model is generated by
\begin{align}\label{eq:Gj}
\hat{G}_j=(-1)^{\hat{n}_j}\hat{\tau}^x_{j-1,j}\hat{\tau}^x_{j,j+1}.
\end{align}
This operator has two eigenvalues, $g_j=\pm1$. It is easy to check that $[\hat{H}_0,\hat{G}_j]=0,\,\forall j$, embodying the gauge invariance of the system. 

A building block of this model has recently been implemented using a Floquet setup in an ultracold-atom experiment \cite{Schweizer2019}. The implementation is based on periodically driving the two-component ultracold atoms in a double-well potential resonantly at the on-site interaction strength \cite{Barbiero2019}. With the appropriate modulation parameters, the resulting effective Floquet Hamiltonian hosts a $\mathbb{Z}_2$ gauge symmetry. Nevertheless, the driving also leads to terms that explicitly break the $\mathbb{Z}_2$ gauge symmetry. These errors have been quantified to take the form \cite{Schweizer2019}
\begin{align}\nonumber
	\lambda\hat{H}_1=\,\lambda\sum_{j=1}^{L-1}\Big[&\big(\alpha_1\hat{a}_j^\dagger\hat{\tau}^+_{j,j+1} \hat{a}_{j+1}+\alpha_2\hat{a}_j^\dagger \hat{\tau}^-_{j,j+1} \hat{a}_{j+1}+\mathrm{H.c.}\big)\\\label{eq:H1}
	&+\big(\alpha_3\hat{n}_j-\alpha_4\hat{n}_{j+1}\big)\hat{\tau}^z_{j,j+1}\Big],
\end{align}
on an extended lattice, where the coefficients $\alpha_{1,\ldots,4}$ depend on the driving parameters of the Floquet setup, and are normalized such that they sum to unity. At small $\lambda$, without any protection term, gauge invariance in this model exhibits intrinsic timescales $\propto J^{s-1}/\lambda^{s},\,s=1,2,\ldots,L/2$ over which it persists in a staircase of prethermal plateaus in finite systems \cite{Halimeh2020c,Halimeh2020d}. Eventually, however, the system relaxes into a steady state of maximal gauge violation if no gauge protection is employed \cite{Halimeh2020a}.

Gauge protection in the $\mathbb{Z}_2$ LGT has been first addressed using a protection term with the full generator, $\propto V\sum_j\hat{G}_j$ \cite{Halimeh2020a}. Even though the performance has been demonstrated to be formidable, with a controlled-error regime where the violation plateaus at a value $\propto\lambda^2/V^2$ at sufficiently large $V$, implementing such a protection term is quite challenging in current QSM setups, given that $\hat{G}_j$ is a three-body two-species term, see Eq.~\eqref{eq:Gj}. Thereafter, the concept of LPG protection~\eqref{eq:LPGprotection} has been introduced and shown to successfully enable experimentally feasible gauge-symmetry stabilization, with the LPG, given by
\begin{align}\label{eq:LPG}
\hat{W}_j=\hat{\tau}^x_{j-1,j}\hat{\tau}^x_{j,j+1}+2g_j^\text{tar}\hat{n}_j,
\end{align}
composed of at most two-body single-species terms \cite{Halimeh2021stabilizing}. These are experimentally easier to implement than the ideal gauge theory~\eqref{eq:H0} itself. 

However, the analysis in Ref.~\cite{Halimeh2021stabilizing} has only focused on finite systems and perturbative errors. Even though its analytic derivations guarantee that LPG protection works in the thermodynamic limit with a compliant sequence $c_j$, i.e., one that satisfies condition~\eqref{eq:compliance}, the latter will grow exponentially in system size, rendering it impractical for large systems. Furthermore, an experimentally feasible sequence in the thermodynamic limit, such as the noncompliant periodic sequence $c_j=[6(-1)^j+5]/11$ used in Ref.~\cite{Halimeh2021stabilizing} for finite systems, has been analytically predicted to completely fail in the thermodynamic limit in a worst-case scenario. 

In the following, we will demonstrate numerically, and argue analytically, why LPG protection far outperforms this scenario. For this purpose, we prepare our system in a gauge-invariant initial state $\ket{\psi_0}$ in the target sector $\hat{G}_j\ket{\psi_0}=\ket{\psi_0}$ (i.e., $g_j^\text{tar}=1$), $\forall j$, and then quench with the faulty theory $\hat{H}=\hat{H}_0+\lambda\hat{H}_1+V\hat{H}_W$. We are interested in the resulting quench dynamics of the temporally averaged gauge violation, the staggered boson occupation, and the total electric flux,
\begin{subequations}
\begin{align}\label{eq:viol}
\varepsilon(t)&=1-\frac{1}{Lt}\int_0^t ds \sum_j\bra{\psi(s)}\hat{G}_j\ket{\psi(s)},\\\label{eq:matter}
n_\text{stag}(t)&=\frac{1}{L} \sum_j(-1)^j\bra{\psi(t)}\hat{n}_j\ket{\psi(t)},\\\label{eq:electric}
E(t)&=\frac{1}{L} \sum_j\bra{\psi(t)}\hat{\tau}^x_{j,j+1}\ket{\psi(t)},
\end{align}
\end{subequations}
respectively, where $\ket{\psi(t)}=e^{-i\hat{H}t}\ket{\psi_0}$. We will numerically calculate this quench for an analog quantum simulator in Sec.~\ref{sec:analog} and for a quantum computer in Sec.~\ref{sec:digital}.

\begin{figure}[htp]
	\centering
	\includegraphics[width=.48\textwidth]{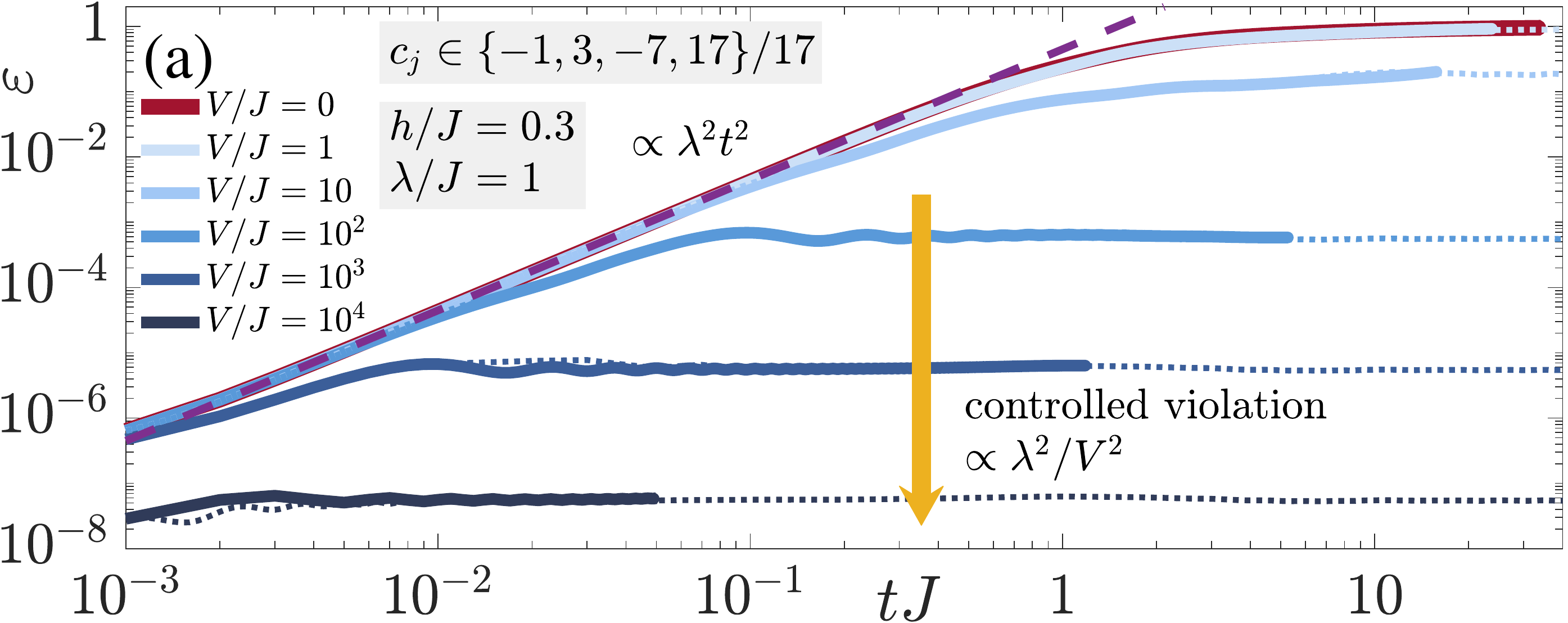}\\
	\vspace{1.1mm}
	\includegraphics[width=.48\textwidth]{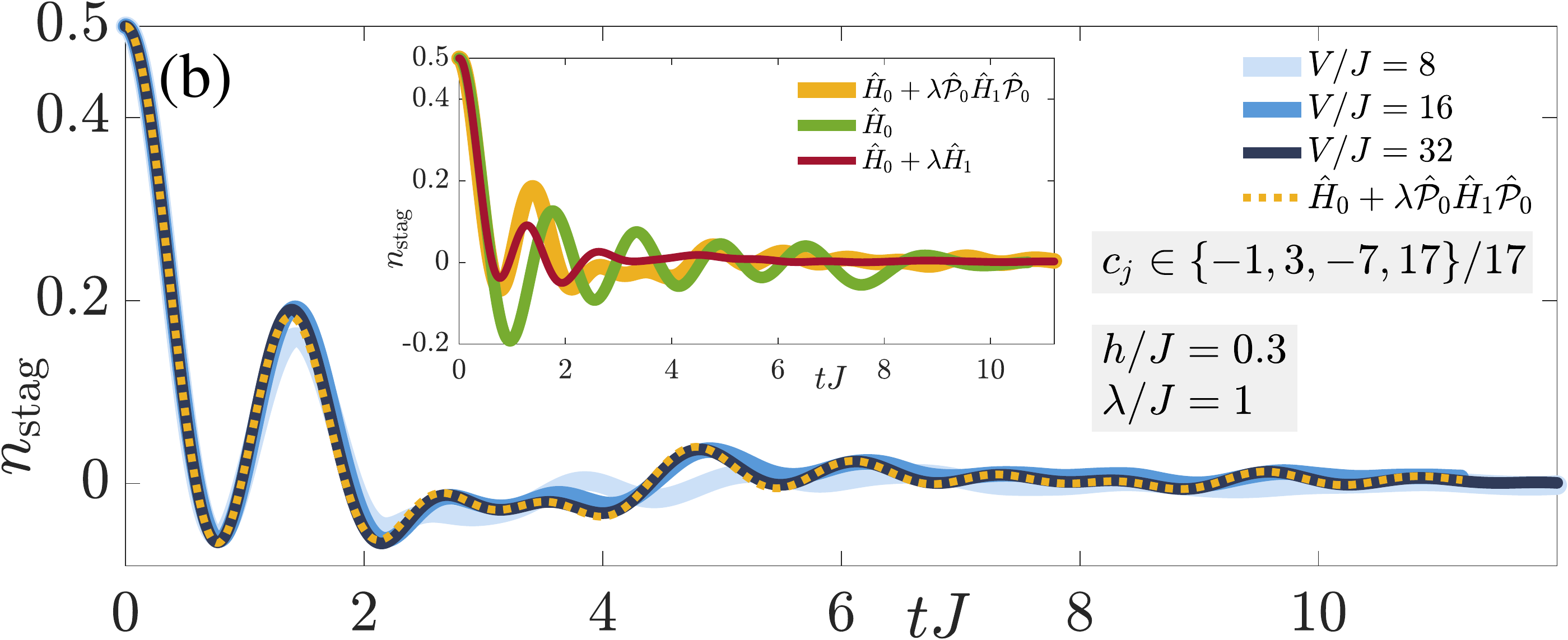}\\
	\vspace{1.1mm}
	\includegraphics[width=.48\textwidth]{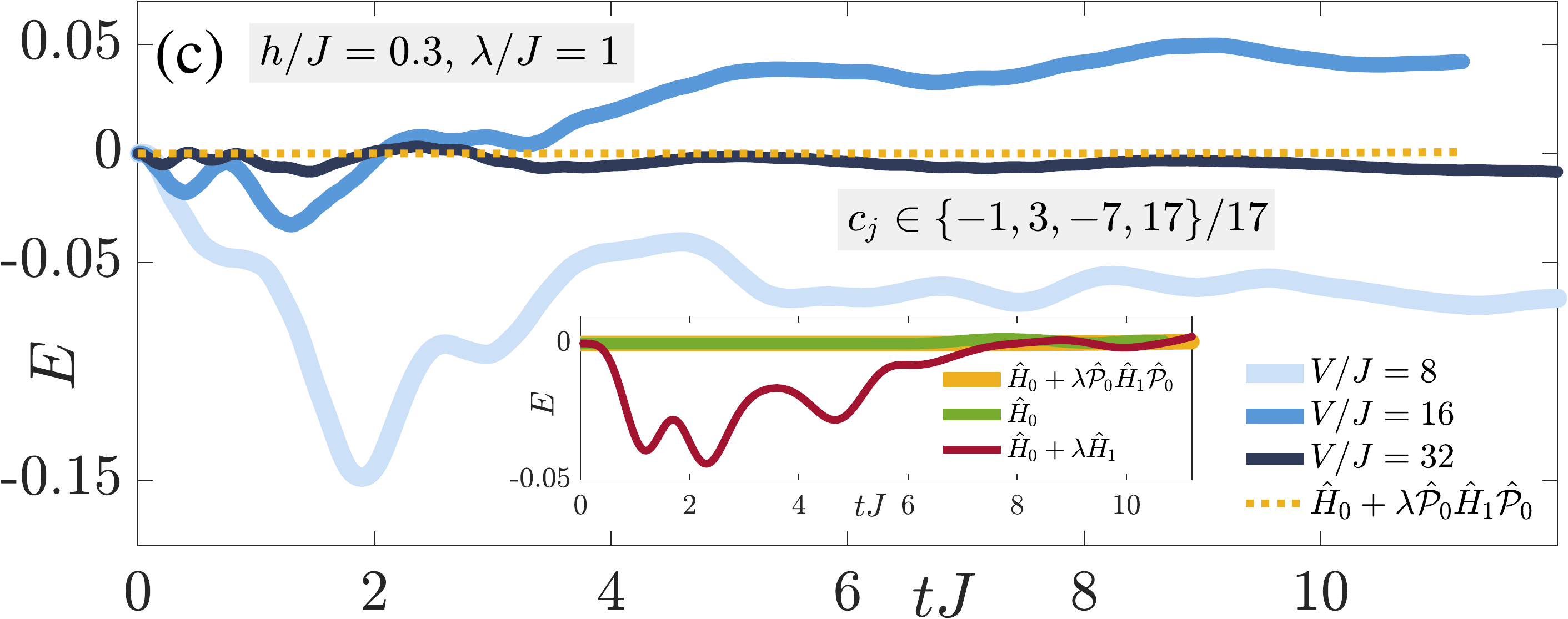}
	\caption{(Color online). Quench dynamics under the faulty theory $\hat{H}=\hat{H}_0+\lambda\hat{H}_1+V\hat{H}_W$ obtained from iMPS in the thermodynamic limit starting in the staggered-matter state defined by the unit cell $\ket{\bullet\rightarrow\circ\rightarrow\bullet\leftarrow\circ\leftarrow}$, which lies in the target sector $\hat{G}_j\ket{\psi_0}=\ket{\psi_0},\,\forall j$. The LPG protection employs the sequence $c_j\in\{-1,3,-7,17\}/17$, which although compliant over the unit cell, is noncompliant in the thermodynamic limit. (a) The gauge violation~\eqref{eq:viol} shows two distinct behaviors over the accessible evolution times in iMPS. For sufficiently small protection strength $V$, the gauge violation will continue growing towards a maximal value in the presence of gauge-breaking errors ($\lambda>0$). At sufficiently large $V$, the gauge violation enters a plateau of value $\propto\lambda^2/V^2$ at a timescale $\propto1/V$, and this plateau seems stable at all evolution times reached in iMPS, despite the errors being nonperturbative in strength ($\lambda=J$). Corresponding exact diagonalization results on a finite system with four matter sites and four links under periodic boundary conditions (dotted lines) show excellent agreement with their counterparts in the thermodynamic limit (solid lines), indicating that the gauge violation, at least over accessible evolution times, is system-size independent under LPG protection. (b,c) As analytically derived using the quantum Zeno effect, the LPG protection leads to the emergence of an adjusted gauge theory $\hat{H}_\text{adj}=\hat{H}_0+\lambda\hat{\mathcal{P}}_0\hat{H}_1\hat{\mathcal{P}}_0$, with $\hat{\mathcal{P}}_0$ the projector onto the target gauge sector. For $V=16J$ and $\lambda=J$, we find that the LPG protection yields dynamics that are very well reproduced by the adjusted gauge theory for the case of the staggered boson number~\eqref{eq:matter} in the thermodynamic limit, therefore surpassing worst-case analytic predictions of Ref.~\cite{Halimeh2021stabilizing}. The total electric flux~\eqref{eq:electric} is also well-reproduced by $\hat{H}_\text{adj}$ at $V=32J$. The insets show that the dynamics at $V=0$ and $\lambda=J$ significantly deviates from that under the adjusted gauge theory. This highlights the strength of the LPG protection in that such nonperturbative errors are efficiently suppressed in the thermodynamic limit at experimentally feasible values of $V$.}
	\label{fig:Neel} 
\end{figure}

\section{Analog quantum simulation in the thermodynamic limit}\label{sec:analog}

The simulation of gauge theories on large-scale quantum simulators has recently garnered a lot of attention, allowing the observation of gauge invariance in a ramp protocol through the Coleman phase transition \cite{Yang2020}, and the investigation of thermalization dynamics \cite{Zhou2021} in the $\mathrm{U}(1)$ QLM through implementations employing energetic protection schemes. The LPG protection is a very promising scheme for the realization of large-scale experimental implementations of the $\mathbb{Z}_2$ LGT with stable gauge invariance \cite{Homeier2021}. It is therefore instructive to investigate how LPG protection will fare in the thermodynamic limit for accessible evolution times. A prominently successful numerical method for this purpose is the infinite matrix product state (iMPS) technique, which we employ here within the framework of the time-dependent variational principle \cite{Haegeman2011,Haegeman2016,Vanderstraeten2019,MPSKit}. For our most stringent numerical simulations, we find that convergence of our iMPS results is achieved with a time-step of $10^{-3}/J$ and a bond dimension of $400$.

Let us first take the initial state to be the staggered-matter state defined as $\ket{\bullet\rightarrow\circ\rightarrow\bullet\leftarrow\circ\leftarrow}$ on a unit cell composed of four matter sites ($\bullet\equiv$ occupied, $\circ\equiv$ empty) and four gauge links (arrows denote eigenstates of local electric-flux operator) that repeats indefinitely (thermodynamic limit). This initial state is in the target sector $\hat{G}_j\ket{\psi_0}=\ket{\psi_0},\,\forall j$ and is at half-filling of the global $\mathrm{U}(1)$ symmetry due to the conservation of boson number in our model. We now quench the staggered-matter state with the faulty-theory Hamiltonian $\hat{H}=\hat{H}_0+\lambda\hat{H}_1+V\hat{H}_W$ defined in Eqs.~\eqref{eq:H0},~\eqref{eq:H1}, and~\eqref{eq:LPGprotection}. The LPG is given in Eq.~\eqref{eq:LPG} with $g_j^\text{tar}=1$, while the employed sequence is $c_j\in\{-1,3,-7,17\}/17$, which is compliant over the four-site four-link unit cell but, since it repeats indefinitely, is a noncompliant sequence in the thermodynamic limit. Furthermore, we choose $\alpha_1=0.5110$, $\alpha_2=-0.4953$, $\alpha_3=0.7696$, and $\alpha_4=0.2147$ in Eq.~\eqref{eq:H1}, which correspond to an experimentally friendly value of the driving parameter in the Floquet setup of Ref.~\cite{Schweizer2019}. However, we stress that this choice of parameters is not at all special otherwise, and we have checked that our qualitative conclusions are qualitatively the same for other values of these coefficients. 

The resulting dynamics of the gauge violation~\eqref{eq:viol} at a fixed nonperturbative error strength $\lambda=J$ are shown for various values of the protection strength $V$ in Fig.~\ref{fig:Neel}(a). The gauge violation grows $\propto\lambda^2t^2$ at early times, as can be calculated from time-dependent perturbation theory (TDPT) \cite{Halimeh2020a}. Thereafter, two distinct regimes are apparent in this plot. The first occurs at sufficiently small $V$, where the gauge violation grows uncontrollably towards a maximal value. The second regime occurs at sufficiently large $V$, whereby the gauge violation stops increasing at a timescale $\propto1/V$ and enters a stable plateau with a value $\propto\lambda^2/V^2$. This plateau has been analytically derived in degenerate perturbation theory in the case of full protection $\propto V\sum_j\hat{G}_j$ \cite{Halimeh2020a}, which has also been analytically predicted to reliably stabilize gauge invariance for a volume-independent value of $V$ \cite{Halimeh2020e,vandamme2021reliability} based on formalism extended from the case of slow heating in fast-driven systems \cite{abanin2017rigorous}. It is therefore quite remarkable that the LPG protection with a noncompliant sequence shows qualitatively similar behavior in the thermodynamic limit, which far exceeds the analytic predictions of Ref.~\cite{Halimeh2021stabilizing}.

In the framework of iMPS, we are limited in terms of maximal evolution times that we can access, and guaranteeing the stability of the violation plateaus at much longer times is not possible. However, it is worthy to note here that there is great agreement in the plateau values at a given value of $V$ between the iMPS result and its counterpart from ED for the corresponding system of four matter sites and four gauge links with periodic boundary conditions, as shown in Fig.~\ref{fig:Neel}(a). In order to rule out trivial or localized dynamics that would make going to larger systems redundant, we show in Appendix~\ref{app:supp} that the dynamics of observables such as the staggered boson occupation is significantly different at finite size from its corresponding counterpart in the thermodynamic limit. Moreover, despite their limited maxima, the times achieved in Fig.~\ref{fig:Neel}(a) are relevant from the perspective of ongoing QSM experiments. For instance, the recent large-scale Bose--Hubbard superlattice implementation of a $\mathrm{U}(1)$ gauge theory \cite{Zhou2021} reaches evolution times of approximately $3/J$. In our iMPS calculations, we reach a maximal evolution time of $t\approx10/J$ in the controlled-error regime of $V=100J$, see Fig.~\ref{fig:Neel}(a). This is well within the range of maximal lifetimes achieved in state-of-the-art QSM platforms.

We now turn our attention to local observables, which in a typical experiment are more easily accessible than the gauge violation, although the latter has recently been observed \cite{Yang2020}. In particular, we show the dynamics of the staggered boson occupation~\eqref{eq:matter} in Fig.~\ref{fig:Neel}(b) for the staggered-matter initial state. Already at $V=16J$, the dynamics under the faulty theory $\hat{H}$ is well-reproduced by the emergent \textit{adjusted gauge theory} $\hat{H}_\text{adj}=\hat{H}_0+\lambda\hat{\mathcal{P}}_0\hat{H}_1\hat{\mathcal{P}}_0$, with $\hat{\mathcal{P}}_0$ the projector onto the target gauge sector, where the derivation of $\hat{H}_\text{adj}$ is based on the quantum Zeno effect \cite{facchi2002quantum,facchi2004unification,facchi2009quantum,burgarth2019generalized}. It is important to note that starting in the target gauge sector, $\hat{H}_\text{adj}$ faithfully reproduces the dynamics of the faulty theory at sufficiently large $V$ within an error upper bound $\propto tV_0^2L^2/V$, with $V_0$ an energy term depending on the parameters of $\hat{H}_0$ but not on $V$ \cite{Halimeh2020e,Halimeh2021stabilizing}. In other words, up to a timescale $\tau_\text{adj}\propto V/(V_0 L)^2$, the dynamics under $\hat{H}$ are guaranteed to exhibit controlled gauge invariance, which is essential for faithfully simulating gauge-theory dynamics in a QSM setup. However, in the thermodynamic limit, this timescale goes to zero, predicting a worst-case scenario where the faulty theory can produce no controlled gauge-theory dynamics. As such, the results of Fig.~\ref{fig:Neel}(b) show that this worst-case scenario can be avoided up to all accessible times using LPG protection even with a noncompliant sequence in the presence of nonperturbative errors. Moreover, $\lambda=J$ and $V=16J$ are well within the feasible parameter ranges that can be accessed in modern QSM setups \cite{Hauke2013,Schweizer2019,Yang2020,Zhou2021}. This is indeed impressive given that here the gauge-breaking error is nonperturbative, yet already $V\lesssim16J$ is sufficient to produce stable gauge-theory dynamics up to $t\gtrsim 10/J$, which are relatively large experimental lifetimes. It is also important to note here that the unprotected dynamics ($V=0$) significantly deviates from that under the adjusted gauge theory (see inset), indicating that the error term is by no means benign. Therefore, LPG protection is crucial in stabilizing the gauge-theory dynamics in the presence of such errors. Turning our attention to the total electric flux~\eqref{eq:electric} in Fig.~\ref{fig:Neel}(c), we arrive at the same qualitative conclusions as in the case of the staggered boson number, albeit here a larger $V\approx 32J$ is required for the total electric flux to agree in its dynamics under $\hat{H}$ with that under $\hat{H}_\text{adj}$. Nevertheless, such a value of the gauge protection strength is still within the range of current experiments, and it is important to note that the electric-flux dynamics in this case seem to be trivial in that they average over time to a value close to zero.

\begin{figure}[t!]
	\centering
	\includegraphics[width=.48\textwidth]{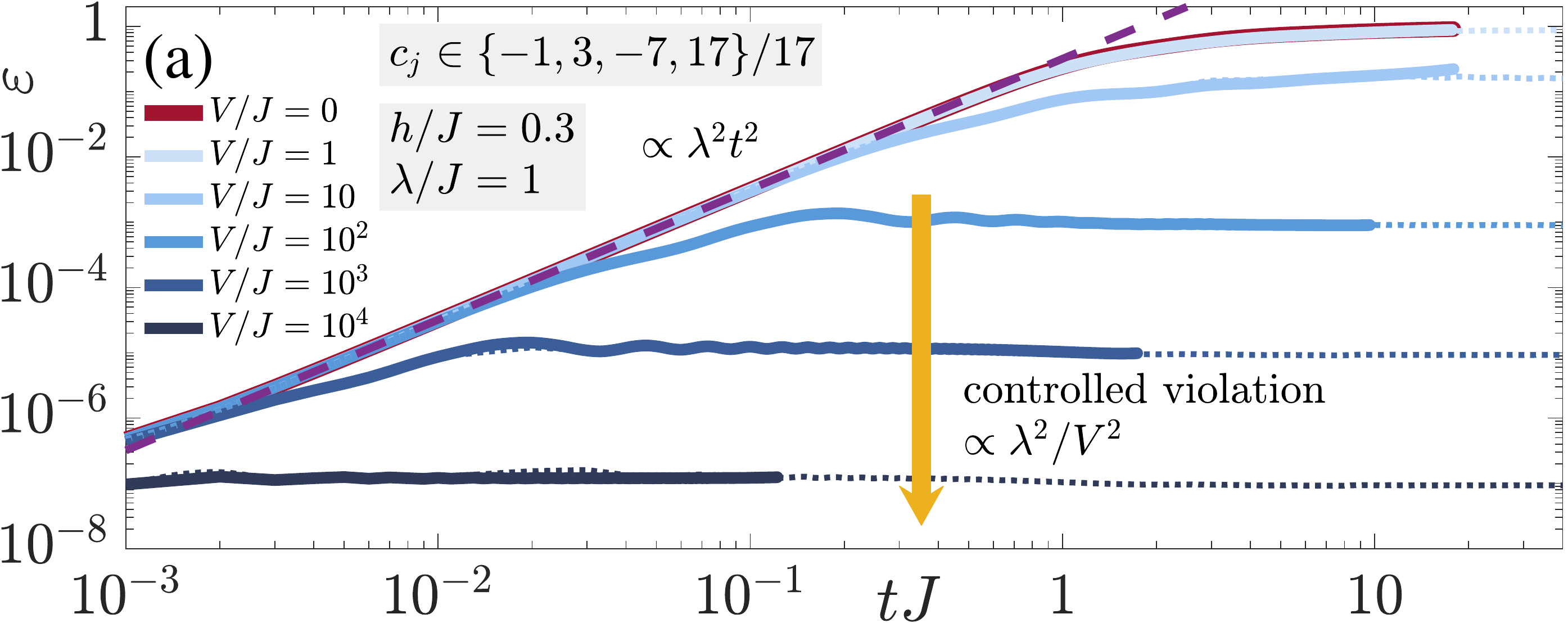}\\
	\vspace{1.1mm}
	\includegraphics[width=.48\textwidth]{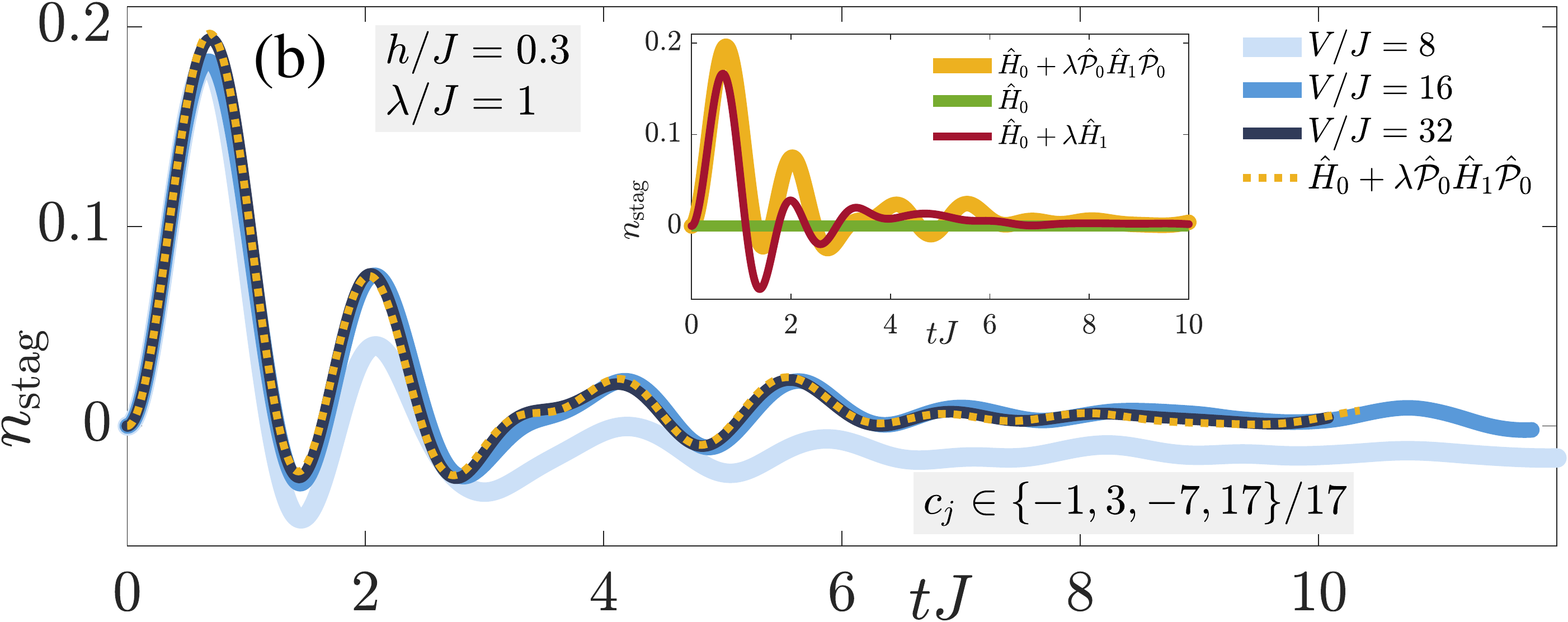}\\
	\vspace{1.1mm}
	\includegraphics[width=.48\textwidth]{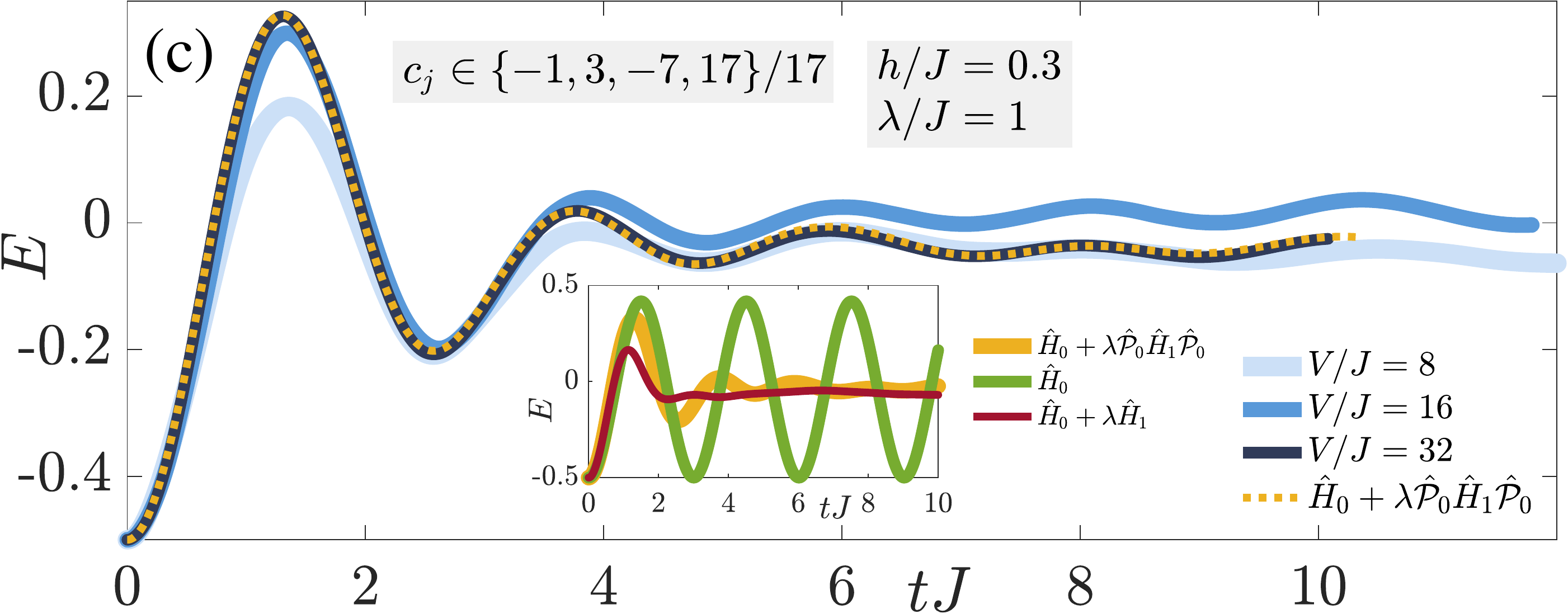}
	\caption{(Color online). Same as Fig.~\ref{fig:QC_Neel} but starting in the charge-density-wave (CDW) state defined by the unit cell $\ket{\bullet\rightarrow\bullet\leftarrow\circ\leftarrow\circ\leftarrow}$, which lies in the target sector $\hat{G}_j\ket{\psi_0}=\ket{\psi_0},\,\forall j$. The qualitative behavior is identical to that of the staggered-matter initial state, indicating the generality of LPG protection with a noncompliant sequence in the thermodynamic limit in the presence of nonperturbative gauge-breaking errors.}
	\label{fig:CDW} 
\end{figure}

In order to check the generality of these results, we now consider the charged-density-wave (CDW) initial state defined as $\ket{\bullet\rightarrow\bullet\leftarrow\circ\leftarrow\circ\leftarrow}$ on an indefinitely repeating four-site four-link unit cell (thermodynamic limit). As shown in Fig.~\ref{fig:CDW}, the conclusions are qualitatively identical to those of the staggered-matter initial state of Fig.~\ref{fig:Neel}, with the electric flux showing even better quantitative agreement with the adjusted-theory prediction, which we attribute to its dynamics being nontrivial compared to the case of the staggered-matter initial state. Indeed, the two-regime picture of the gauge violation is also present in Fig.~\ref{fig:CDW}(a) with $\varepsilon(t)$ rendered controlled at a stable value $\propto\lambda^2/V^2$ after a timescale $\propto1/V$ at sufficiently large $V$, whereas at values of $V$ that are too small, it seems to continually grow in time. Once again, we see that the gauge violation for the corresponding finite system with four matter sites and four gauge links with periodic boundary conditions is essentially quantitatively identical to its counterpart in the thermodynamic limit.

The robustness of the quantum Zeno effect is confirmed in Fig.~\ref{fig:CDW}(b-c): the dynamics of the staggered boson occupation~\eqref{eq:matter} and electric flux~\eqref{eq:electric}, respectively, under the faulty theory show very good agreement with those from the adjusted gauge theory for $V\gtrsim16J$ over all accessible evolution times in iMPS, once again exceeding analytically predicted worst-case scenarios for the thermodynamic limit \cite{Halimeh2021stabilizing}.

Even though we have set $\lambda=J$ and $h=0.3$ for the results of Figs.~\ref{fig:Neel} and~\ref{fig:CDW}, we have checked that our conclusions remain qualitatively the same for different values of these parameters.

As a summary of this section, we have shown that LPG protection with a properly chosen sequence that is compliant over the unit cell, though noncompliant over the whole lattice, will produce reliable gauge invariance with a violation controlled $\propto\lambda^2/V^2$ for all accessible evolution times in iMPS even in the presence of nonperturbative gauge-breaking errors. For moderately small values of the protection strength $V$, we find that the dynamics of local observables is already well-reproduced by an emergent gauge theory during all the evolution times we reach. This shows that LPG protection has substantial potential to stabilize current and future QSM implementations of $\mathbb{Z}_2$ LGTs and possibly other more general gauge theories.

\section{Quench dynamics on a quantum computer}\label{sec:digital}

\begin{figure}[t!]
	\centering
	\includegraphics[width=.48\textwidth]{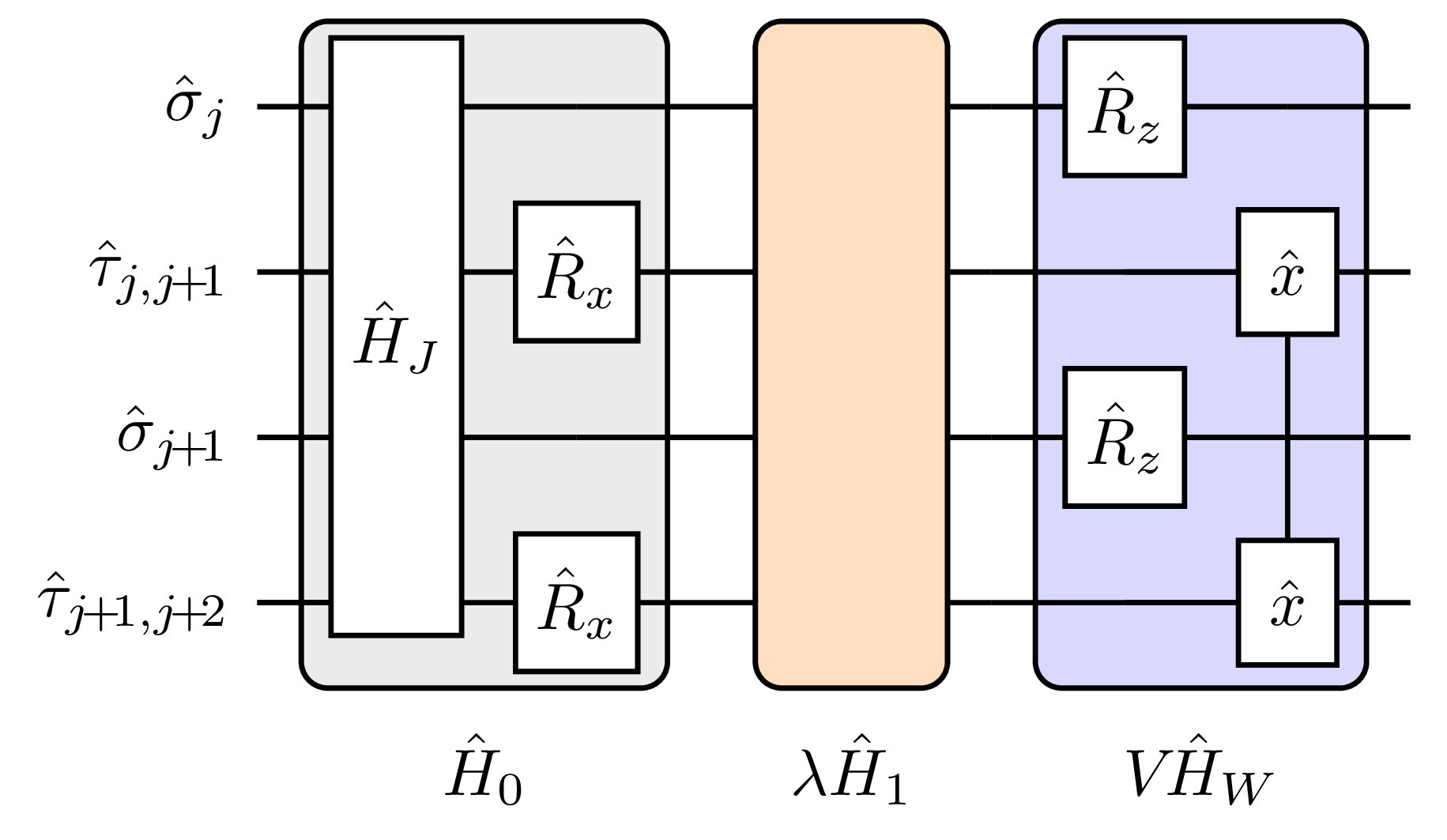}
	\caption{(Color online). Elementary unit of a single Trotter step. Qubits representing matter (gauge) fields are denoted by $\hat{\sigma}$ ($\hat{\tau}$). The implementation of the time evolution by $\hat{H}_0$ of Eq.~\eqref{eq:H0_circuit} is a combination of the layered three-qubit interactions between qubits representing both matter and gauge fields, i.e., $\exp\big(-i\hat{H}_J\delta t\big)$, and single-qubit $x$-rotations $\hat{R}_x$ on the qubits representing gauge fields. In NISQ devices, the ideal dynamics will typically be plagued by gauge-violating error terms $\lambda\hat{H}_1$, which will depend on the implementation. To mitigate them, the LPG protection term $V\hat{H}_W$ of Eq.~\eqref{eq:LPG} can be implemented by single-qubit $z$-rotations $\hat{R}_z$ on matter qubits and the two-qubit gates acting on gauge qubits $\exp(-ic_jV\hat{\tau}_{j,j+1}^x\hat{\tau}_{j+1,j+2}^x\delta t)$, which is denoted by $\hat{x}$-$\hat{x}$.}
	\label{fig:QC_diagram} 
\end{figure}

\begin{figure*}[t!]
	\centering
	\includegraphics[width=.48\textwidth]{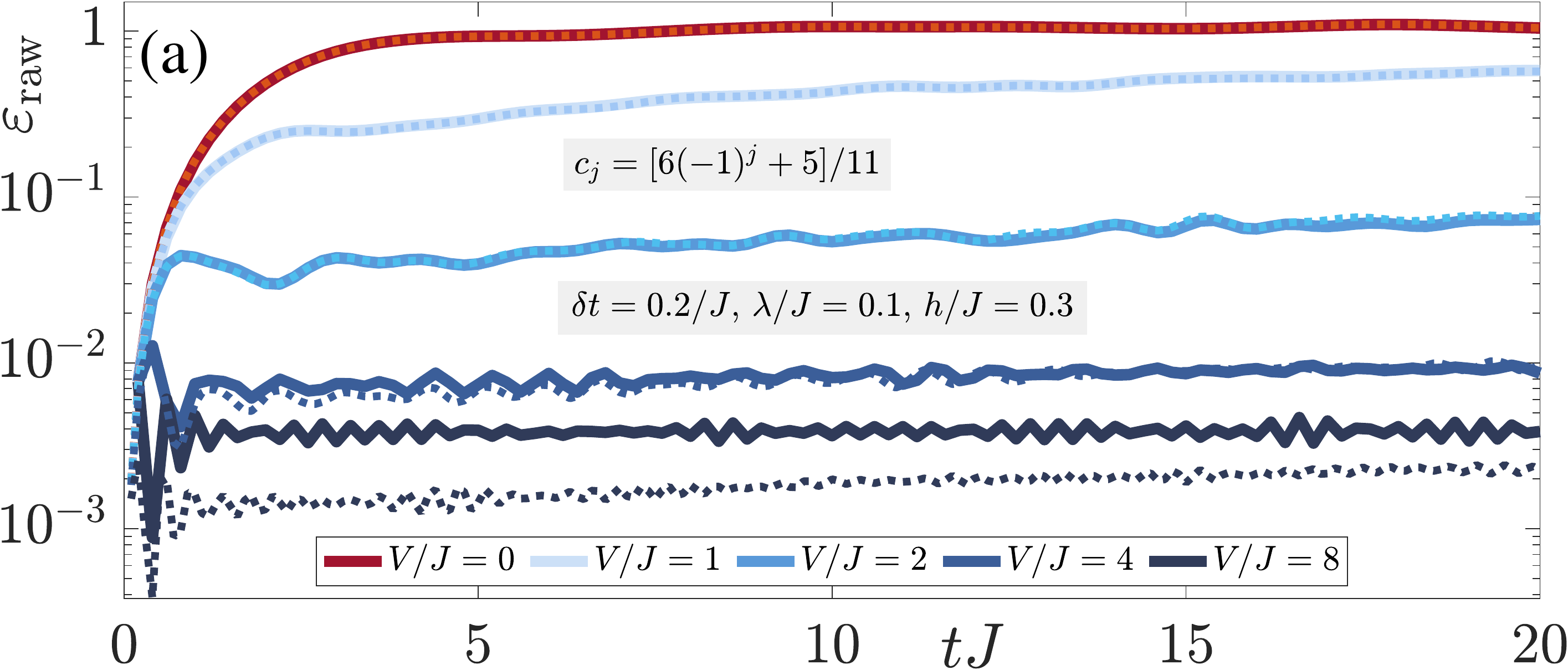}\quad
	\includegraphics[width=.48\textwidth]{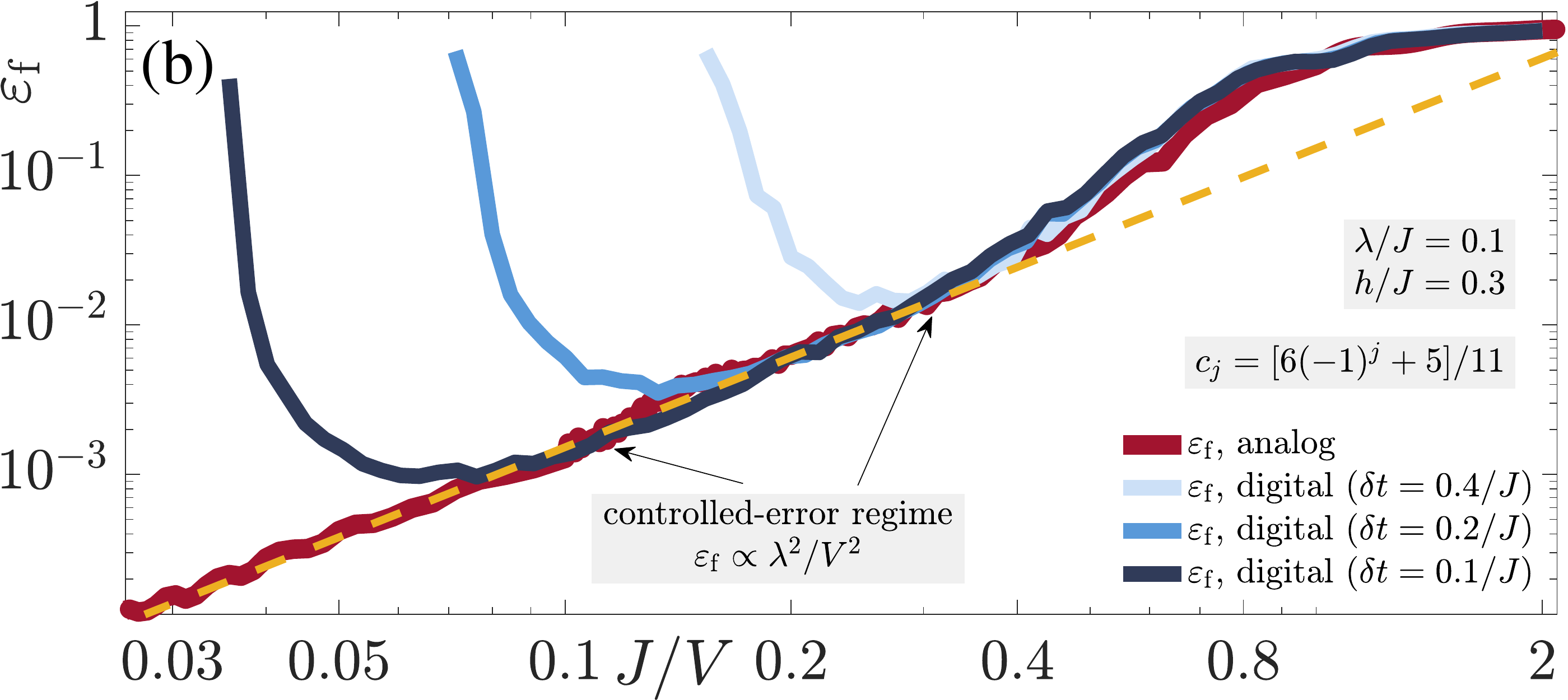}\\
	\vspace{1.1mm}
	\includegraphics[width=.48\textwidth]{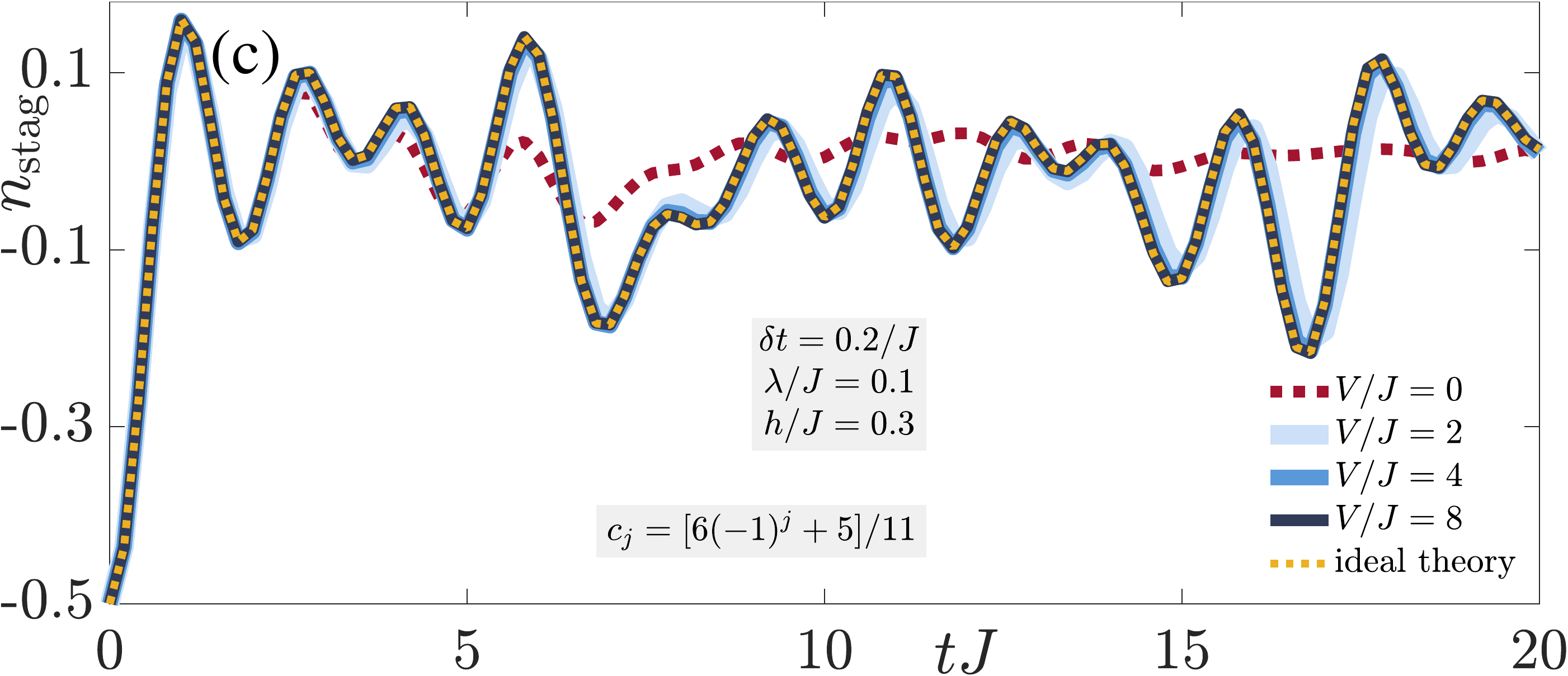}\quad
	\includegraphics[width=.48\textwidth]{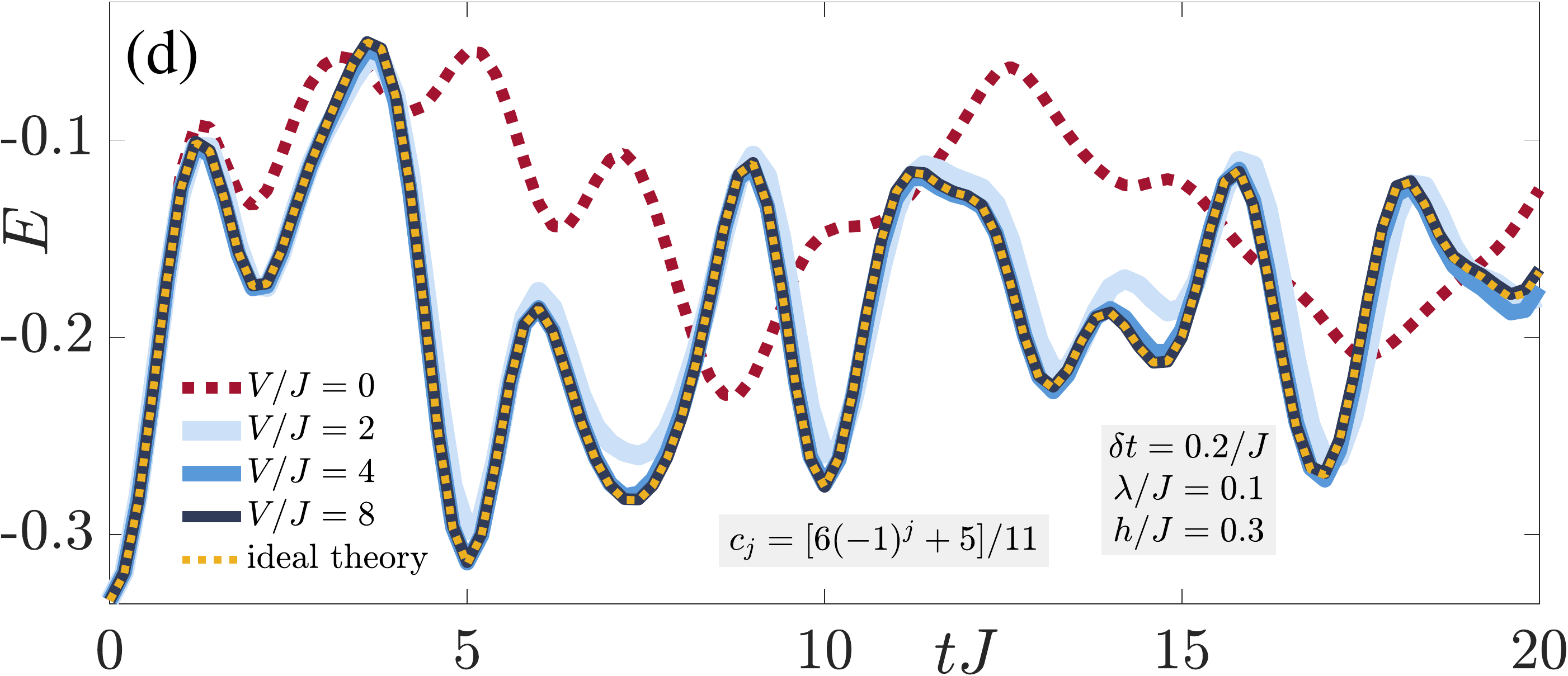}
	\caption{(Color online). Quench dynamics on a finite quantum circuit of the initial staggered-matter state $\ket{\bullet\leftarrow\circ\leftarrow\bullet\rightarrow\circ\rightarrow\bullet\leftarrow\circ\leftarrow}$ under the faulty theory $\hat{H}=\hat{H}_0+\lambda\hat{H}_1+V\hat{H}_W$, simulated from quantum-circuit numerical calculations with open boundary conditions using Cirq. The LPG protection includes the noncompliant sequence $c_j=[6(-1)^j+5]/11$. (a) The time evolution of the raw gauge violation~\eqref{eq:viol_raw} with a Trotter time-step $\delta t=0.2/J$ (solid lines) shows that at moderate values of $V$ the violation plateaus at a timescale $\propto1/V$ to a value $\propto\lambda^2/V^2$ over all calculated evolution times. The corresponding analog results (dotted lines) are shown for comparison. Agreement with the digital result is quite good, except for larger values of $V$ that approach $V_\text{ideal}\sim\pi/(2\delta t)$ (see main text for details). (b) A two-regime picture appears when scanning the value of the gauge violation at the final evolution time $t=20/J$, where for a given Trotter time-step $\delta t$, for sufficiently large $V/\lambda$ but with $V<V_\text{ideal}$, the gauge violation is controlled $\propto\lambda^2/V^2$. Outside this regime, the value of the gauge violation cannot be directly inferred from the value of $V$. Without any protection, (c) the staggered boson number~\eqref{eq:matter} and (d) the total electric flux~\eqref{eq:electric} quickly and significantly deviate from the ideal-theory dynamics, which for the case of the error term~\eqref{eq:H1_circuit} is the ``adjusted'' gauge theory obtained from the quantum Zeno effect. Upon introducing LPG protection, the ideal-theory dynamics is restored already at moderate values of $V$, allowing the use of a larger Trotter time-step, thereby reducing computational overhead on a quantum computer. Even though here we have set $h=0.3J$, we have checked that our results are qualitatively the same for different values of $h$.}
	\label{fig:QC_Neel} 
\end{figure*}

We now investigate the efficacy of LPG protection by numerically simulating a digital quantum computer that implements the gauge-theory dynamics through discrete, Trotterized time-steps. In recent years, quantum advantage beyond classical capabilities has been demonstrated on various quantum computing platforms such as superconducting quantum devices and photonic systems \cite{Arute2019,Zhong2020,wu2021strong,zhong2021phaseprogrammable}, and the associated rapid engineering progress has motivated the use of such devices to probe high-energy physics phenomena theoretically and experimentally \cite{Byrnes2006,Pasquans_review,Yao2020,Brower2020,dejong2021quantum,Buser2021,klco2021standard,alam2021quantum,cohen2021quantum,funcke2021quantum}.

In this vein, we again consider the $(1+1)-$dimensional $\mathbb{Z}_2$ LGT, but rewrite its Hamiltonian as
\begin{align}\label{eq:H0_circuit}
	\hat{H}_0=-J\sum_{j=1}^{L-1}\big(\hat{\sigma}^+_j\hat{\tau}^z_{j,j+1}\hat{\sigma}^-_{j+1}+\text{H.c.}\big)-h\sum_{j=1}^L\hat{\tau}^x_{j,j+1},
\end{align}
where we have now replaced the hard-core boson notation with its equivalent representation in Pauli matrices, as is the traditional convention when dealing with quantum circuits. We consider a one-dimensional arrangement of qubits that alternatingly represent matter ($\hat{\sigma}$) and gauge ($\hat{\tau}$) fields, and we employ open boundary conditions; cf.~Fig.~\ref{fig:QC_diagram}.

\begin{figure*}[t!]
	\centering
	\includegraphics[width=.48\textwidth]{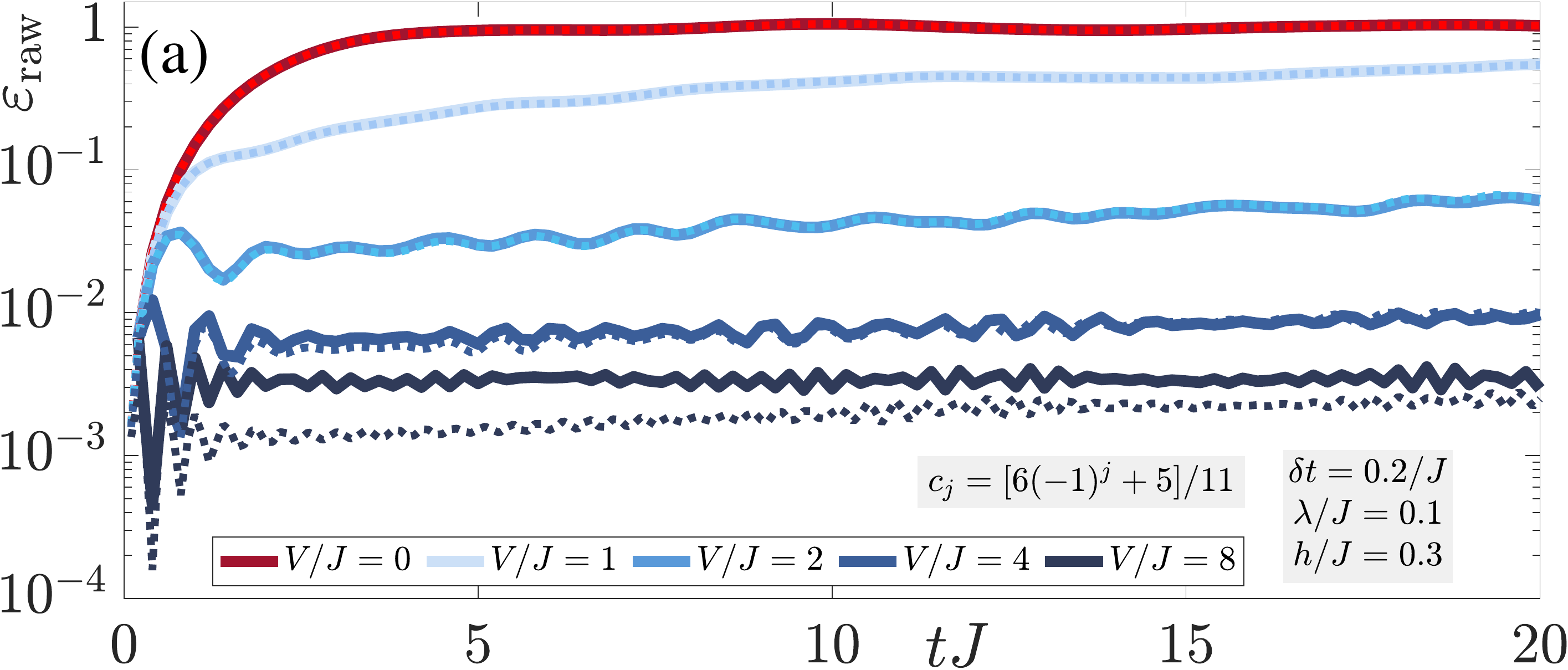}\quad
	\includegraphics[width=.48\textwidth]{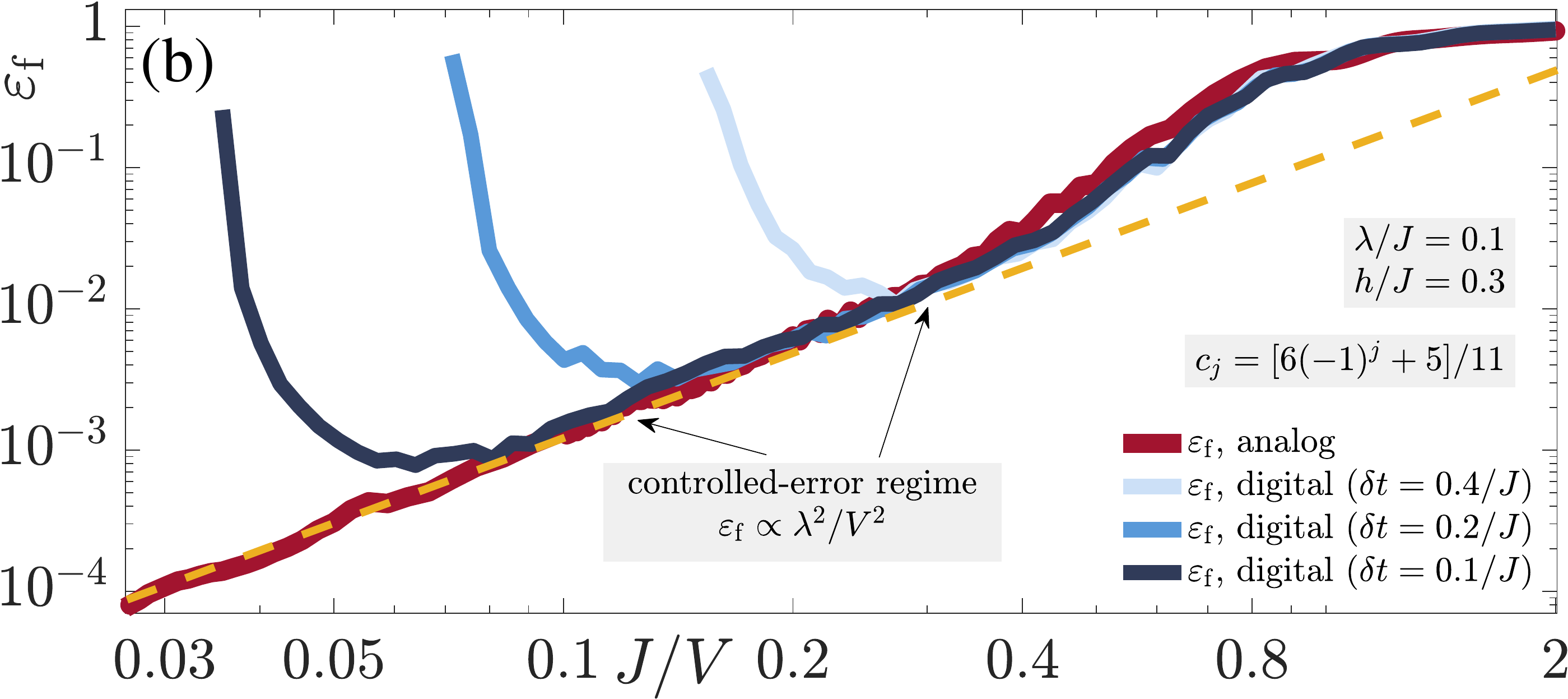}\\
	\vspace{1.1mm}
	\includegraphics[width=.48\textwidth]{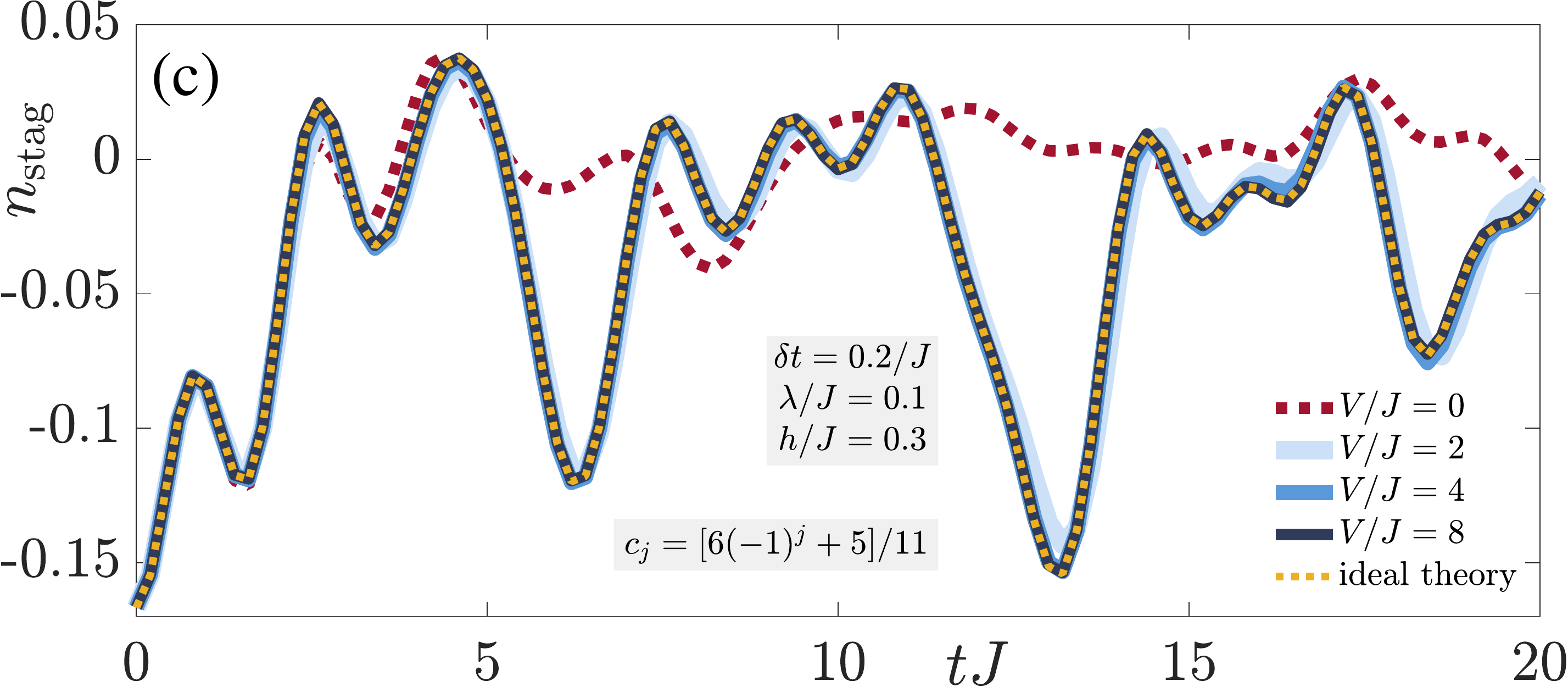}\quad
	\includegraphics[width=.48\textwidth]{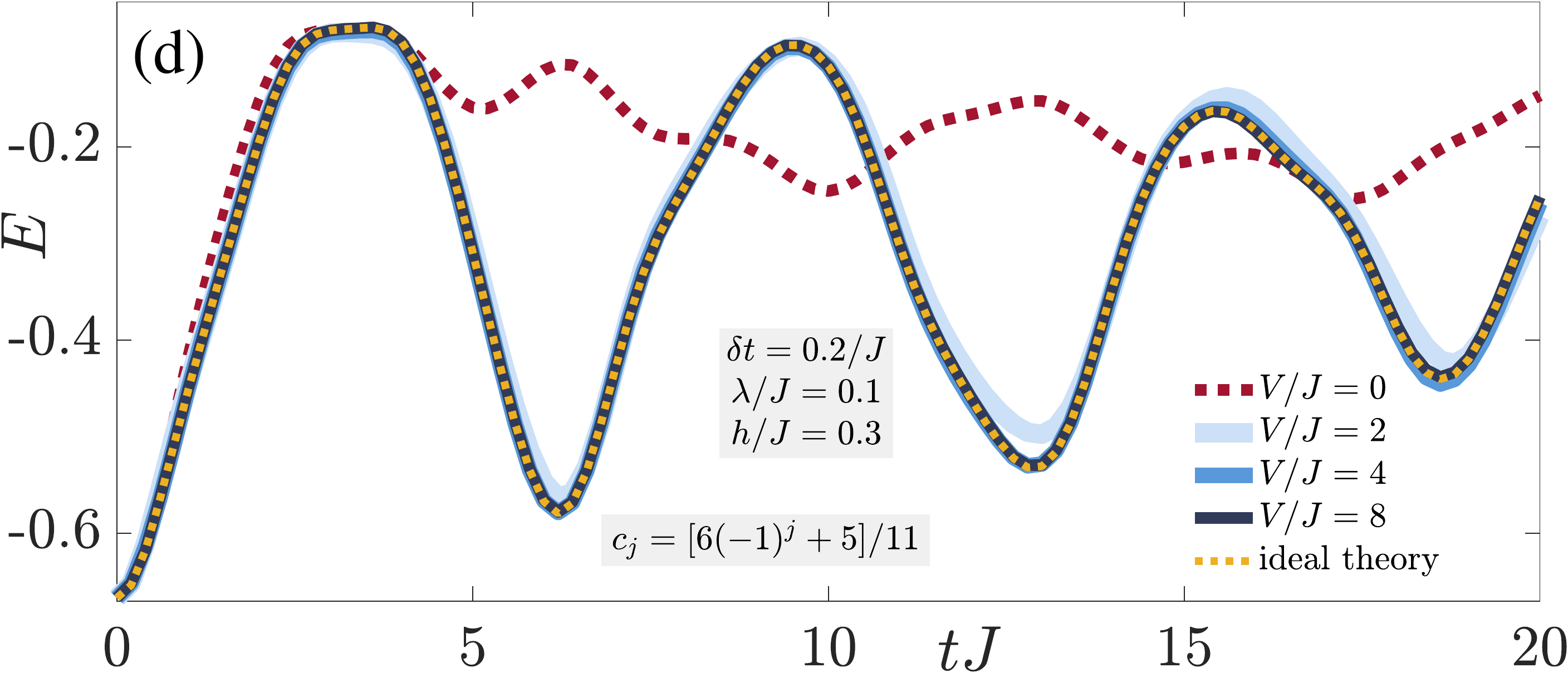}
	\caption{(Color online). Same as Fig.~\ref{fig:QC_Neel} but starting in the domain-wall initial state $\ket{\bullet\leftarrow\bullet\rightarrow\bullet\leftarrow\circ\leftarrow\circ\leftarrow\circ\leftarrow}$. The qualitative conclusions are identical to those in the case of the staggered-matter initial state, showing that the efficacy of LPG protection is general and not restricted to particular fine-tuned initial conditions.}
	\label{fig:QC_DW} 
\end{figure*}

Near-term realizations on quantum computers in the NISQ era will unavoidably be subject to coherent errors, e.g., due to imperfect gate calibrations or further systematic errors. In order to mimic these, we exemplarily choose as explicitly gauge-violating terms both phase flips on gauge fields and unassisted tunneling between matter sites, described by the term
\begin{align}\label{eq:H1_circuit}
	\lambda\hat{H}_1 = \lambda\sum_j (\hat{\tau}^z_{j,j+1}+\hat{\sigma}^+_j \hat{\sigma}^-_{j+1}+\text{H.c.}),
\end{align}
where, in following with implementational relevance \cite{preskill2021quantum}, we will set $\lambda=0.1J$ for the rest of our discussion. To constrain the gauge violation introduced by $\lambda\hat{H}_1$, we make use of the protection term $V\hat{H}_W$ of Eq.~\eqref{eq:LPGprotection} where the LPG is given in Eq.~\eqref{eq:LPG} with $\hat{n}_j=(\hat{\sigma}^z_j+\mathds{1})/2$. 

The time evolution propagated by the faulty theory $\hat{H}=\hat{H}_0+\lambda\hat{H}_1+V\hat{H}_W$ is now implemented in layers $\text{exp}(-i\hat{H}_J\delta t)$, $\text{exp}(-i\hat{H}_h\delta t)$, $\text{exp}(-i\lambda\hat{H}_1\delta t)$, and $\text{exp}(-iV\hat{H}_W\delta t)$ with the Trotter time-step $\delta t$. To this end, we separate the $\mathbb{Z}_2$ LGT Hamiltonian $\hat{H}_0$ into the three-body terms coupling matter and gauge fields, $\hat{H}_J=-J\sum_{j=1}^{L-1}(\hat{\sigma}^+_j\hat{\tau}^z_{j,j+1}\hat{\sigma}^-_{j+1}+\text{H.c.})$, and $\hat{H}_h=-h\sum_{j=1}^L\hat{\tau}^x_{j,j+1}$. The Trotterized time evolution by $\hat{H}_h$ and the matter-site term in $V\hat{H}_W$, i.e., $V\sum_jc_jg_j^{\mathrm{tar}}\hat{\sigma}_j^z$, are implemented by single-qubit rotations $\hat{R}_{x,(j,j+1)}(\phi)=\exp(-i\hat{\tau}^x_{j,j+1}\phi/2)$ and $\hat{R}_{z,j}(\phi)=\exp(-i\hat{\sigma}^z_j\phi/2)$, where the rotation angle $\phi$ is given by the relative weight in the Hamiltonian, e.g., $\exp\big(-i\hat{H}_h\delta t\big)=\bigotimes_j\exp(i\hat{\tau}^x_{j,j+1}h\delta t)=\bigotimes_j\hat{R}_{x,(j,j+1)}(-2h\delta t)$. The implementation of $\exp\big(-i\hat{H}_J\delta t\big)$ consists of layered three-qubit interactions, which we assume to be ideal here as we focus on the effects and suppression of explicit gauge violations introduced by $\lambda \hat{H}_1$. The remainder of the LPG protection term $V\hat{H}_W$, i.e., the part acting on the gauge-fields, can then be implemented by two-qubit gates as $\exp(-ic_jV\hat{\tau}_{j,j+1}^x\hat{\tau}_{j+1,j+2}^x\delta t)$. With this, the implementation of the LPG protection term is less complex than that of the lattice gauge theory itself, which is beneficial for experimental realizations. A sketch of the resulting circuit model is given in Fig.~\ref{fig:QC_diagram}.

We choose an alternating, noncompliant sequence of coefficients $c_j=[6(-1)^j+5]/11$, illustrating that no fine-tuning of parameters is required. Taking the system size to be $2L=12$ sites ($L=6$ matter sites and $L=6$ gauge links), we consider as initial states $\ket{\psi_0}$ the staggered-matter state $\ket{\bullet\leftarrow\circ\leftarrow\bullet\rightarrow\circ\rightarrow\bullet\leftarrow\circ\leftarrow}$ as well as the domain-wall state $\ket{\bullet\leftarrow\bullet\rightarrow\bullet\leftarrow\circ\leftarrow\circ\leftarrow\circ\leftarrow}$. These lie in the target sector $\hat{G}_j\ket{\psi_0}=\ket{\psi_0}$, i.e., $g_j^\text{tar}=1,\,\forall j$ (to define the constraint at $j=1$ in the presence of open boundary conditions, one can add a fictitious electric field at $j-1,1$ that is fixed to the value $+1$). To perform these calculations, we employ Cirq \cite{cirq}, a Python library dedicated to designing and optimizing quantum circuits.

Focusing first on the case of the staggered-matter initial state, we present the corresponding quench dynamics of the raw (i.e., not temporally averaged) gauge violation 
\begin{align}\label{eq:viol_raw}
\varepsilon_\mathrm{raw}(t)=1-\frac{1}{L-1}\sum_{j>1}\bra{\psi(t)}\hat{G}_j\ket{\psi(t)},
\end{align} 
its final value $\varepsilon_\mathrm{f}=\varepsilon_\mathrm{raw}(t=20/J)$, the staggered boson number~\eqref{eq:matter}, and the total electric flux~\eqref{eq:electric}, respectively, in Fig.~\ref{fig:QC_Neel}. These results are shown for a Trotter time-step $\delta t=0.2/J$ with up to $100$ Trotter steps, which covers abundantly the lifetimes achievable with current quantum processors \cite{preskill2021quantum}.

The gauge violation shows two distinct regimes in Fig.~\ref{fig:QC_Neel}(a) over the accessible evolution times. For very small values of the protection strength $V$, the violation seems uncontrolled and increases steadily throughout the time evolution. At moderate values of $V$, the gauge violation is suppressed $\propto\lambda^2/V^2$, just like on an analog quantum simulator. Indeed, we show in the same plot the corresponding analog results (i.e., continuous time; dotted lines), and the agreement with their digital counterparts is very good. The behavior is further confirmed when scanning the final value of the gauge violation at $t=20/J$ as a function of $J/V$, shown in Fig.~\ref{fig:QC_Neel}(b). The transition from an uncontrolled-error regime to a controlled one where the violation scales $\propto\lambda^2/V^2$ occurs around $V\approx2.5J$ when $\lambda=0.1J$ for all Trotter time-steps considered (see legend). Going to smaller Trotter time-steps further extends the controlled-error regime, but then the maximal evolution times will be more challenging to reach in a realistic device with finite gate fidelity. 

Indeed, there is an ideal choice of the protection strength $V_\text{ideal}$ for any given Trotter time-step $\delta t$. As the existence of such a $V_\text{ideal}$ is a consequence of the periodic redundancy of gate angles in the circuit model, it is roughly given as $V_\text{ideal}\sim\pi/(2\delta t)$. Larger choices of $V>V_\text{ideal}$ therefore induce less effective protection and result in a higher $\varepsilon_\mathrm{f}$. For sufficiently large $V/\lambda$ with $V$ also sufficiently smaller than $V_\text{ideal}$, LPG protection displays the controlled-error regime where the gauge violation is suppressed $\propto(\lambda/V)^2$ and lies on top of the corresponding analog result (thick red line). Thanks to the comparatively small values of $V$ required to be in the controlled-error regime, one is able to use relatively large Trotter time-steps, which reduces computational time, and hence incoherent errors, on an actual quantum computer. This highlights a very useful facet of LPG protection on quantum computers.

We now turn our attention to the staggered boson number $n_\mathrm{stag}(t)$ and the total electric flux $E(t)$ in Fig.~\ref{fig:QC_Neel}(c,d), respectively, in order to see how well LPG protection restores the ideal-theory dynamics. As discussed in Sec.~\ref{sec:analog} and Ref.~\cite{Halimeh2021stabilizing}, in the presence of an error term $\lambda \hat{H}_1$, LPG protection at sufficiently large volume-independent protection strength $V$ gives rise to an adjusted gauge theory $\hat{H}_\mathrm{adj}=\hat{H}_0+\lambda\hat{\mathcal{P}}_0\hat{H}_1\hat{\mathcal{P}}_0$. Analytically, this is derived from the quantum Zeno effect \cite{facchi2002quantum,facchi2004unification,facchi2009quantum,burgarth2019generalized}. Unlike for the error term of Eq.~\eqref{eq:H1}, $\hat{\mathcal{P}}_0\hat{H}_1\hat{\mathcal{P}}_0=0$ for the error term of Eq.~\eqref{eq:H1_circuit}, rendering the adjusted gauge theory identical to the ideal theory $\hat{H}_0$. We see that the dynamics under the ideal theory is essentially fully restored for a moderate protection-strength value of $V=4J$ within the accessible evolution times calculated in Cirq for both $n_\mathrm{stag}(t)$ and $E(t)$. This is quite impressive considering that the unprotected faulty theory $\hat{H}_0+\lambda\hat{H}_1$ quickly and significantly deviates in its dynamics from the ideal gauge theory $\hat{H}_0$.

In order to show that our conclusions are general, we now repeat these results for the domain-wall initial state in Fig.~\ref{fig:QC_DW}. The conclusions are qualitatively identical to those obtained in the case of the staggered-matter initial state. The two-regime behavior of the gauge violation is clear, where at or above moderate values of $V$ the gauge violation is suppressed $\propto\lambda^2/V^2$ over all evolution times $t\gtrsim1/V$, as shown in Fig.~\ref{fig:QC_DW}(a,b). Once again, the quantum Zeno effect demonstrates its power in Fig.~\ref{fig:QC_DW}(c,d), with the LPG protection restoring the ideal-theory dynamics at moderate values of $V$ for both the staggered boson number and the total electric flux.

\section{Conclusion and outlook}\label{sec:conc}
We have performed numerical calculations using the infinite matrix product state technique, which works directly in the thermodynamic limit, of quench dynamics in gauge theories with experimentally relevant nonperturbative gauge-breaking errors mitigated by local-pseudogenerator protection. At experimentally feasible values of the protection strength, we find that LPG protection with an experimentally friendly noncompliant sequence in the thermodynamic limit stabilizes gauge invariance up to all accessible evolution times, which translate to state-of-the-art lifetimes in modern experimental platforms. Furthermore, LPG protection leads to the emergence of an adjusted gauge theory due to the quantum Zeno effect that faithfully reproduces the dynamics for all accessible times at moderate values of the protection strength. With the current pursuit of large-scale analog quantum simulations of lattice gauge theories in various quantum synthetic matter setups, our results show that an experimentally feasible protection scheme has impressive potential even in the thermodynamic limit, surpassing expectations analytically predicted in Ref.~\cite{Halimeh2021stabilizing}.

Furthermore, we have showcased the performance of LPG protection on finite-size quantum circuits using the quantum circuit toolkit Cirq. It is demonstrated that for typical errors on quantum circuits, relatively small values of the protection strength allow for a controlled suppression of gauge violations up to all relevant lifetimes that can be accessed in modern NISQ devices. Not only that, but LPG protection in this case also reproduces the ideal-theory dynamics of local observables very reliably, despite these errors leading to vastly different dynamics in the unprotected case. Indeed, the small protection strengths at which the LPG protection stabilizes gauge violations allow for using larger Trotter time-steps, thereby reducing computational effort for NISQ devices, which is highly desired in order to minimize effects due to decoherence.

\begin{acknowledgments}
The authors are grateful to Monika Aidelsburger, Annabelle Bohrdt, Lukas Homeier, Haifeng Lang, and Christian Schweizer for collaboration on related projects. This work is part of and supported by Provincia Autonoma di Trento, the ERC Starting Grant StrEnQTh (project ID 804305), the Google Research Scholar Award ProGauge, and Q@TN — Quantum Science and Technology in Trento, Research Foundation Flanders (G0E1520N, G0E1820N), and ERC grants QUTE (647905) and ERQUAF (715861). This research was funded by the Deutsche Forschungsgemeinschaft (DFG, German Research Foundation) under Germany's Excellence Strategy -- EXC-2111 -- 390814868 and via DFG Research Unit FOR 2414 under project number 277974659.
\end{acknowledgments}

\appendix
\section{Supporting numerical results for analog quantum simulation}\label{app:supp}
\subsection{Finite-size results}
\begin{figure}[t!]
	\centering
	\includegraphics[width=.48\textwidth]{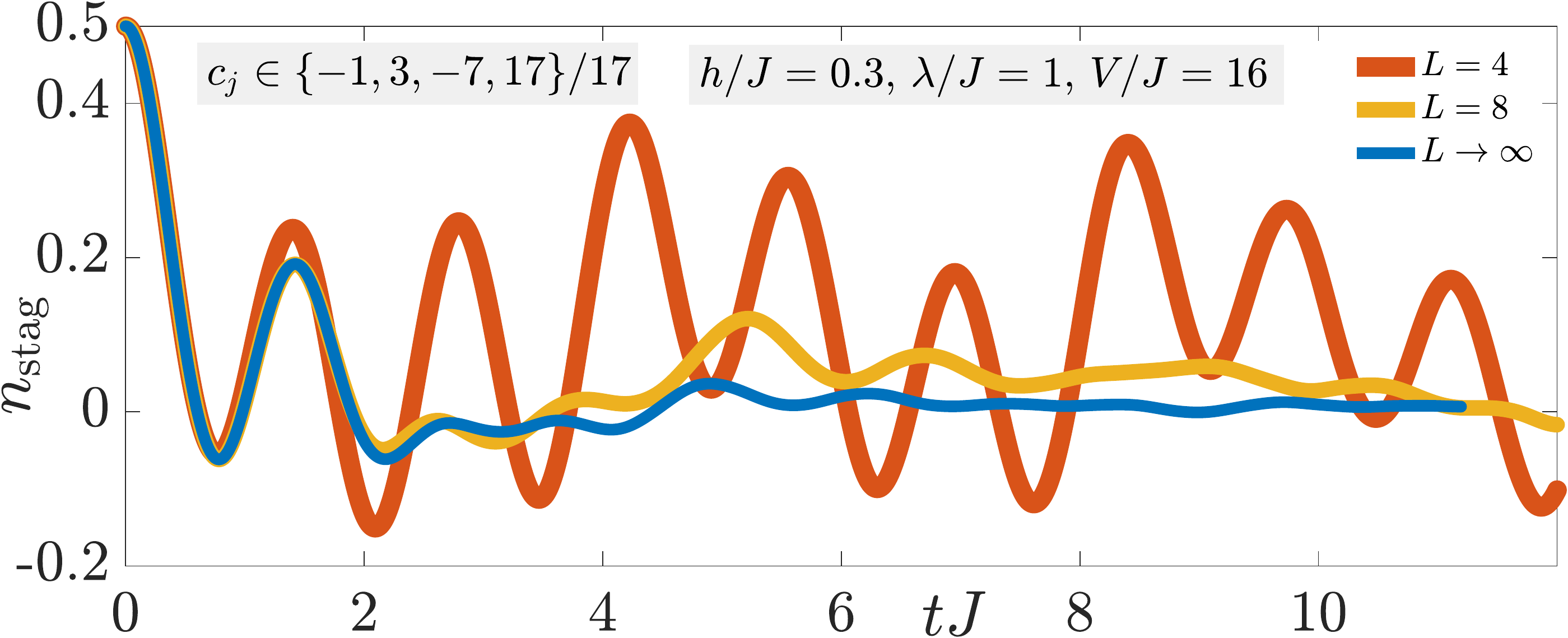}
	\caption{(Color online). Quench dynamics of the staggered boson number~\eqref{eq:matter} under the faulty gauge theory $\hat{H}=\hat{H}_0+\lambda\hat{H}_1+V\hat{H}_W$ with $c_j\in\{-1,3,-7,17\}/17$, and starting in the initial state defined by the unit cell $\ket{\bullet\rightarrow\circ\rightarrow\bullet\leftarrow\circ\leftarrow}$ for the finite sizes $L=4$ and $8$ matter sites (ED+Lanczos calculations, periodic boundary conditions employed), and in the thermodynamic limit (iMPS calculations).}
	\label{fig:FSS} 
\end{figure}

In the main text, we have seen that under the faulty gauge theory $\hat{H}=\hat{H}_0+\lambda\hat{H}_1+V\hat{H}_W$ with $c_j\in\{-1,3,-7,17\}/17$ there is very good qualitative and quantitative agreement in the quench dynamics  of the gauge violation~\eqref{eq:viol} for a finite system of $L=4$ matter sites with periodic boundary conditions and for a system in the thermodynamic limit at $\lambda=J$ and at all values of $V$. However, in Fig.~\ref{fig:FSS}, we see that for $V=16J$ the quench dynamics of the staggered boson number~\eqref{eq:matter} is significantly different between a finite system of $L=4$ matter sites and an infinite system ($L\to\infty$). This is also the case for other values of $V$. As such, the dynamics in the thermodynamic limit is not trivially reproduced in a finite system, which is all the more impressive that LPG protection performs so well in the thermodynamic limit.

\subsection{Choice of noncompliant sequence}

\begin{figure}[t!]
	\centering
	\includegraphics[width=.48\textwidth]{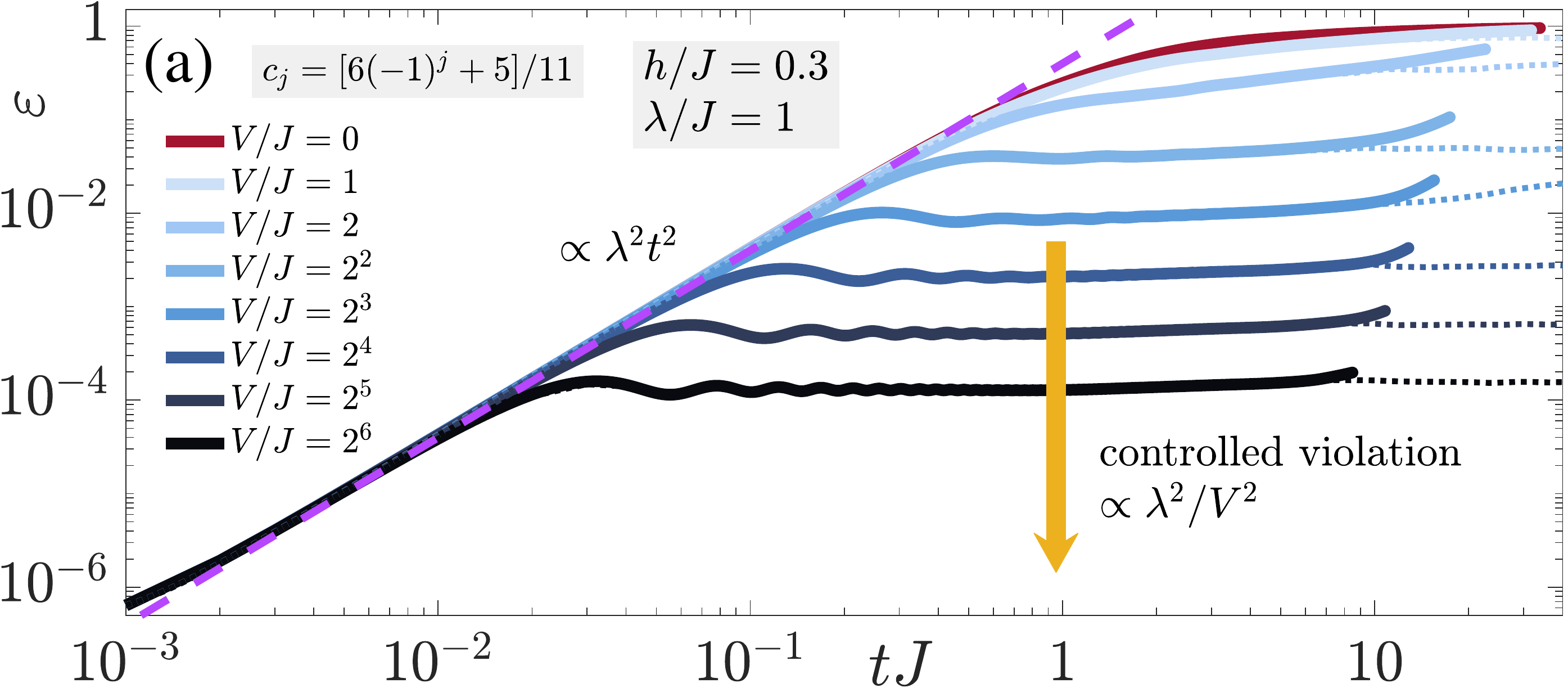}\\
	\vspace{1.1mm}
	\includegraphics[width=.48\textwidth]{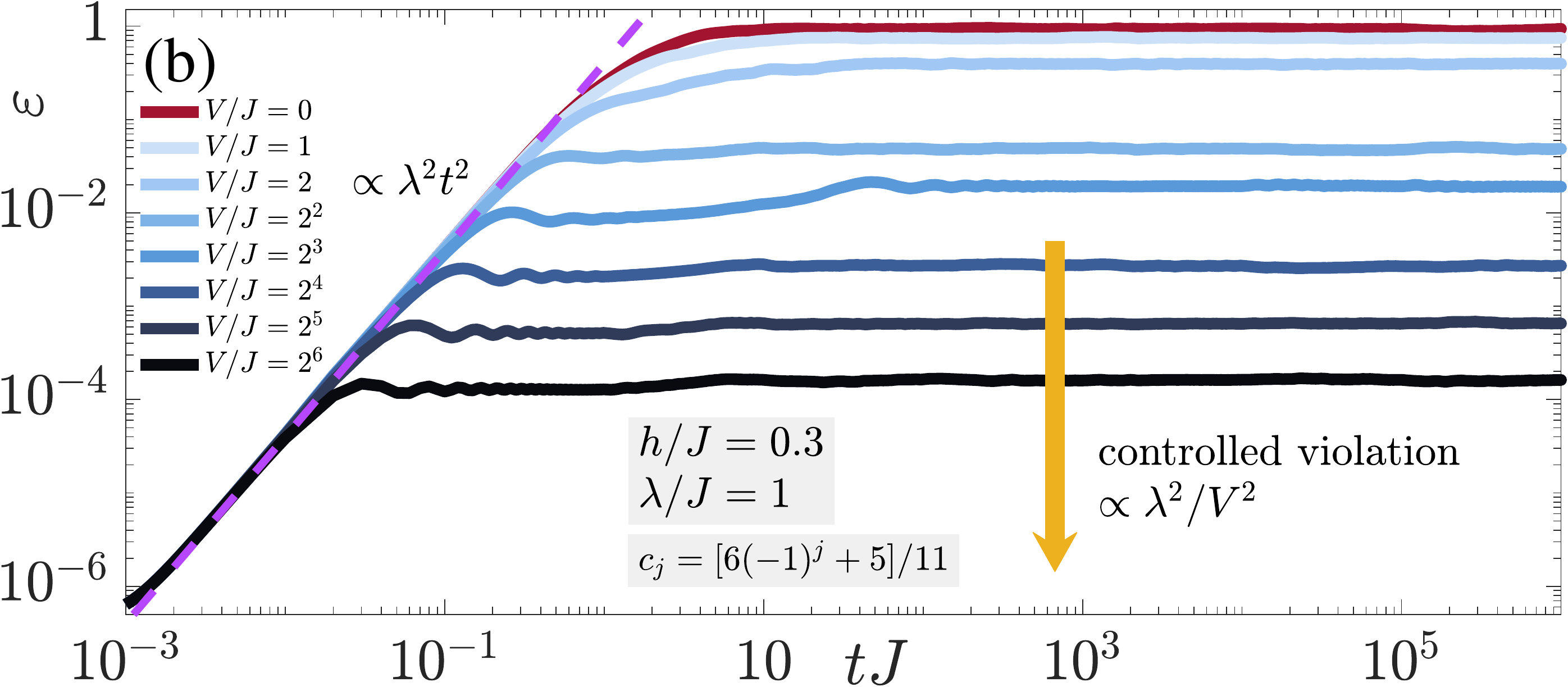}
	\caption{(Color online). Quench dynamics of the gauge violation~\eqref{eq:viol} under LPG protection with the noncompliant sequence $c_j=[6(-1)^j+5]/11$ in (a) the thermodynamic and (b) for a finite system with $L=4$ matter sites and $L=4$ gauge links with periodic boundary conditions. The unit cell defining the initial state is $\ket{\bullet\rightarrow\circ\rightarrow\bullet\leftarrow\circ\leftarrow}$. The gauge violation~\eqref{eq:viol} leaves its early-time growth $\propto\lambda^2t^2$ at a timescale $\propto1/V$, after which it settles into a controlled-violation plateau $\propto\lambda^2/V^2$ at sufficiently large $V$. However, in the thermodynamic limit it then begins to increase slowly at intermediate evolution times and diverges from the finite-system result (dotted lines) at the largest evolution times we can reach in iMPS, with this divergence time being proportional to $V$. In contrast, for the finite system, the gauge violation is controlled at all accessible times for sufficiently large $V$.}
	\label{fig:cj} 
\end{figure}

A compliant sequence over the whole lattice satisfies the condition $\sum_{j=1}^L c_j(w_j-g_j^\text{tar})=0\iff w_j=g_j^\text{tar},\,\forall j$, where $w_j$ are the eigenvalues of the LPGs $\hat{W}_j$. Such a sequence can be analytically shown to protect gauge invariance up to timescales exponential in a volume-independent protection strength $V$ \cite{Halimeh2021stabilizing} through an extension of formalism on slow heating in fast-driven models \cite{abanin2017rigorous}. However, as explained in the main text, this sequence is not feasible for large-scale systems since it grows exponentially with system size. This is the main reason why we employ noncompliant sequences $c_j$ in our iMPS simulations that work directly in the thermodynamic limit. In fact, we would not be able to use iMPS for a sequence $c_j$ that is compliant in the thermodynamic limit, since then the MPS unit cell will have to be infinite in size. Naturally, it is also of experimental relevance to employ noncompliant sequences that repeat in space. In the main text, we have focused in iMPS on the noncompliant sequence $c_j\in\{-1,3,-7,17\}/17$, which is compliant on a finite system of $L=4$ matter sites and $L=4$ gauge links. As shown in Fig.~\ref{fig:Neel}(a), this sequence allows the suppression of gauge violations $\propto\lambda^2/V^2$ for all accessible evolution times in the thermodynamic limit even when gauge-breaking errors are nonperturbative. Furthermore, the gauge violation in the thermodynamic limit agrees quantitatively very well with its counterpart for a finite system with $L=4$ matter sites and $L=4$ gauge links with periodic boundary conditions.

\begin{figure}[t!]
	\centering
	\includegraphics[width=.48\textwidth]{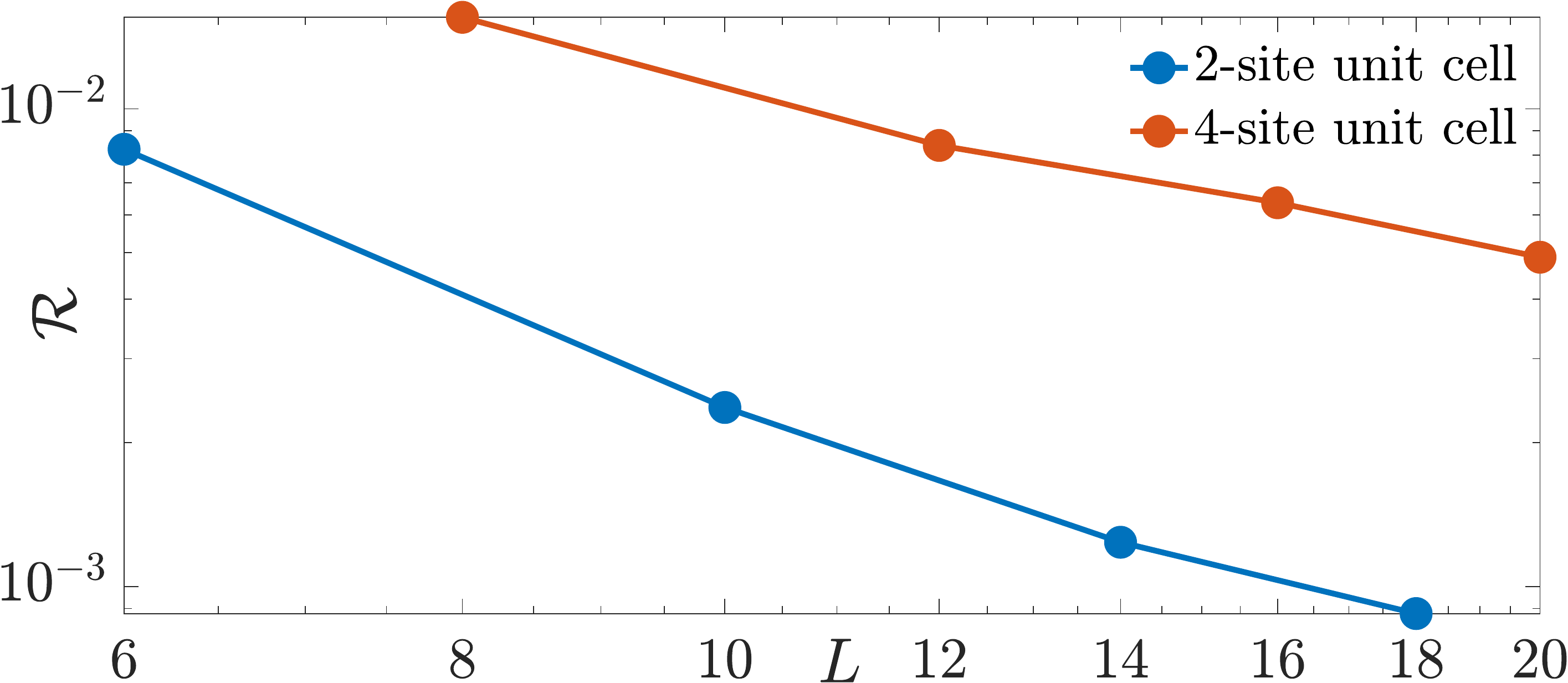}
	\caption{(Color online). The fraction $\mathcal{R}$ of gauge-invariant sectors satisfying $\sum_jc_j(w_j-g_j^\text{tar})=0$ but with $w_j\neq g_j^\text{tar}$ for at least one value of $j$. It steadily decreases with system size in an almost power-law fashion. This may lead to a slowing down of the spread of gauge violations, thereby partly explaining why LPG protection with a noncompliant sequence fares well in the thermodynamic limit.}
	\label{fig:R} 
\end{figure}

It is now interesting to see how the noncompliant sequence $c_j=[6(-1)^j+5]/11$, prominently used in Ref.~\cite{Halimeh2021stabilizing} for perturbative errors on finite systems, will fare in the thermodynamic limit with nonperturbative errors. The corresponding quench dynamics of the gauge violation starting in the initial state defined over the unit cell $\ket{\bullet\rightarrow\circ\rightarrow\bullet\leftarrow\circ\leftarrow}$ are presented in Fig.~\ref{fig:cj}(a). After the expected initial growth $\propto\lambda^2t^2$ at early times, the gauge violation settles into a plateau $\propto\lambda^2/V^2$ at a timescale $\propto1/V$ when $V$ is sufficiently large. However, soon thereafter the gauge violation starts increasing slowly before rapidly diverging away from the corresponding finite-size result (dotted-line) at the largest accessible evolution times in iMPS.  This divergence also seems to occur earlier with larger $V$. If we look at the corresponding finite-size results from ED in Fig.~\ref{fig:cj}(b), we find that at sufficiently large $V$ the gauge violation exhibits two plateaus both of which have values proportional to $\lambda^2/V^2$, but with the first one slightly lower in value, and ending earlier with larger $V$, while the second plateau persists up to all evolution times calculated in ED. This may partly explain the divergence from the ED result in Fig.~\ref{fig:cj}(a), but not completely. In reality, the sequence $c_j$ guarantees controlled suppression of first-order processes in $\hat{H}_1$, since these break two adjacent local constraints, while $c_j\in\{-1,3,-7,17\}/17$ guarantees controlled suppression of up to second-order processes in $\hat{H}_1$. It is therefore intuitive why the latter sequence will perform better than the former. Of course, and as mentioned in the main text, we cannot guarantee that the sequence $c_j\in\{-1,3,-7,17\}/17$ will reliably suppress gauge-breaking errors at times longer than those we can access in iMPS. However, and as we have argued in the main text, the accessible evolution times we reach in iMPS are equivalent to experimental lifetimes in state-of-the-art QSM setups. Therefore, LPG protection is a powerful tool for stabilizing gauge invariance in such experiments.

Finally, let us consider the fraction $\mathcal{R}$ of gauge-invariant sectors resonant with the target gauge sector for a given LPG sequence. These are the gauge-invariant sectors satisfying the relation $\sum_jc_j(w_j-g_j^\text{tar})=0$ but with $w_j\neq g_j^\text{tar}$ for at least one value of $j$. It is interesting to see how $\mathcal{R}$ behaves as a function of system size $L$. As such, we consider the sequence $c_j\in\{-1,3,-7,17\}/17$ and calculate the fraction $\mathcal{R}$ as a function of $L$ in Fig.~\ref{fig:R}, where we see that it actually decreases with system size. This is encouraging news because it means that even though the number of gauge-invariant sectors resonant with the target gauge sector increases with system size, its fraction of the total number of gauge-invariant sectors decreases, and this will slow the proliferation of gauge violations.

\bibliography{LPGthermo_biblio}
\end{document}